\documentclass[
reprint,
superscriptaddress,
amsmath,amssymb,
aps,
prb,
twocolumn
]{revtex4-2}

\usepackage{graphicx}
\usepackage{dcolumn}
\usepackage{bm}
\usepackage[breaklinks]{hyperref}
\hypersetup{colorlinks=true, linkcolor=blue, citecolor=blue, filecolor=blue, urlcolor=blue}
\graphicspath{{figures}}
\usepackage{bm}

\usepackage{graphicx,subfigure}
\usepackage{epsfig}
\usepackage{bm}
\usepackage{dcolumn}
\usepackage{color} 
\usepackage{physics}
\usepackage{float}
\makeatletter
\newcommand*{\rom}[1]{\expandafter\@slowromancap\romannumeral #1@}
\makeatother
\usepackage{lipsum}
\usepackage{subfigure}
\usepackage{dsfont}
\usepackage{enumitem}
\usepackage{tikz}

\usepackage[normalem]{ulem}
\newcommand{\rz}[1]{\textcolor[rgb]{0,0.0,0}{#1}}
\newcommand{\CH}[1]{\textcolor[rgb]{1,0.32,1}{#1}}

\begin{document}
\title{Observation of feedback-directed quantum dynamics in large-scale quantum processors}

\author{Ruizhe Shen}
\email{e0554228@u.nus.edu}
\affiliation{Department of Physics, National University of Singapore, Singapore 117551}

\author{Ching Hua Lee}
\email{phylch@nus.edu.sg}
\affiliation{Department of Physics, National University of Singapore, Singapore 117551}
\date{\today}

\begin{abstract}
Programmable quantum hardware provides an emerging platform for exploring and controlling non-unitary quantum dynamics through measurement-based operations. In this work, we introduce feedback-directed circuit architectures that integrate spatially structured mid-circuit measurements with real-time conditional operations to steer the evolution of random dynamics, and perform their large-scale simulations (up to 100 qubits) on programmable digital quantum processors. By promoting measurement from a passive readout to an active control signal, these adaptive monitored circuits enable directional information flow and generate intrinsic asymmetry in random circuit simulations. We implement this framework on IBM superconducting quantum processors and observe robust, noise-resilient signatures of feedback-induced asymmetry distinct from the more well-known non-Hermitian skin effect. Our results establish feedback as a programmable resource for non-unitary control, opening new avenues for engineering measurement-based dynamics, non-equilibrium phenomena, and tunable open-system behavior on large-scale quantum hardware.
\end{abstract}

\maketitle
\section{Introduction}\label{sec0}
Quantum computers are rapidly evolving into versatile experimental platforms capable of engineering and probing complex quantum dynamics at large scales\cite{preskill2018quantum,kim2023evidence,chen2022high,chen2023robust,miessen2024benchmarking,andersen2025thermalization,guo2024site}. Recent progress in device coherence, control precision, and qubit connectivity has enabled quantum processors to implement long-depth circuits, high-fidelity entangling operations, and mid-circuit measurements, establishing a new frontier for simulating non-unitary physics on programmable architectures~\cite{jiang2025advancements,evered2023high,iqbal2024topological,wu2025measurement,niu2024ac}. In these settings, various non-unitary phenomena~\cite{kawabata2019symmetry,el2018non,ashida2020non,lin2023topological,zhang2022review,poddubny2023interaction,shen2023proposal,yang2025non,liu2025non,xue2026topologically,zhang2021acoustic,cheng2025stochasticity,qin2023non,zhang2022universal,matsumoto2020continuous,lee2020many,li2023many,okuma2020topological,shen2022non,liang2022observation,lee2021many,kawabata2022entanglement,shen2025non,yang2025beyond,okuma2020topological,qin2024kinked,gliozzi2024many}, previously associated with open systems under dissipation or decoherence, can now be robustly emulated and controlled \cite{shen2025observation,shen2024enhanced,dogra2021quantum,shen2025robust,koh2025interacting,zhang2025observation,cao2023probing,lin2022experimental}. At the heart of this emerging direction is the realization that quantum measurements are a powerful source of controllable non-unitarity. Each projective measurement collapses the wavefunction, generating stochastic yet tunable evolution that can profoundly reshape entanglement and information flow. When applied repeatedly or selectively during circuit evolution, such measurements can induce collective dynamical effects, such as measurement-induced phase transitions, crossovers between volume-law and area-law entanglement, and other non-unitary critical behavior~\cite{hines2025pauli,koh2023measurement,deist2022mid,li2018quantum,skinner2019measurement,gullans2020dynamical,gullans2020scalable}.

The feasibility of such measurement-driven protocols has been dramatically accelerated by rapid advances in quantum computing hardware. Among the most transformative advances is the implementation of mid-circuit measurement and feedback-based operations, which enable real-time access to quantum information during circuit execution~\cite{hines2025pauli,koh2024readout,koh2023measurement,deist2022mid,zhu2023interactive,decross2023qubit}. These capabilities support adaptive decision-making and measurement-conditioned operations. This technological progress has given rise to a new generation of hybrid quantum–classical control protocols, in which measurement outcomes are processed by a classical controller that determines the next quantum operation in real time~\cite{baumer2024efficient,gupta2024probabilistic,baumer2024quantum}. Consequently, mid-circuit control has emerged as a practical route to embedding non-unitary evolution within programmable quantum circuits, bridging theoretical concepts of non-unitary dynamics with experimentally realizable platforms.

{Conventionally, measurements have been associated with stochastic wavefunction collapse. But here, we demonstrate that mid-circuit measurements, when spatially structured and combined with feedback-based control, can additionally} generate a new class of feedback-directed quantum dynamics~\cite{friedman2023measurement,nahum2018operator,fisher2023random,pai2019localization,ravindranath2023entanglement,iadecola2023measurement}. In our setup, feedback-based circuits implement an explicitly record-dependent quantum channel, in which the measurement record is promoted to an active control signal that encodes classical information into the quantum evolution. As such, the arrow of information flow within the circuit can be dynamically engineered and controlled, rather than passively inferred.  In this sense, feedback transforms measurement backaction into programmable directionality. More broadly, by coupling the quantum state to its classical measurement record, feedback introduces an operational form of spacetime nonlocality: future operations are conditioned on past outcomes, thereby correlating distant events across the circuit spacetime. When embedded within chaotic (random) unitary circuits, these feedback-based primitives provide a robust and scalable mechanism for steering the evolution of local observables, establishing feedback as a practical resource for engineering directionality and controllable dynamics on quantum hardware.


We validate our feedback-directed quantum dynamical framework through large-scale digital quantum simulations performed on state-of-the-art superconducting devices on the IBM Quantum platform, using up to 100 qubits, a scale far beyond previous attempts in related directions~\cite{koh2023measurement,hothem2025measuring,noel2022measurement}. Despite the inherent noise in current hardware, we observe robust signatures of feedback-directed scrambling: information propagation and local observables exhibit a clear directional asymmetry induced by conditional measurement and control. These results introduce a powerful new paradigm for controlling the flow of quantum information through measurement and feedback, demonstrating that non-unitary quantum circuits can be precisely designed and implemented on present-day quantum hardware. Beyond establishing a route to programmable open-system dynamics, our findings open the door to experimental studies of measurement-induced criticality, nonequilibrium phase transitions, and directional quantum transport in large-scale quantum systems.

\section{{Results}}

\subsection{{Quantum circuits for feedback-based control}}


In the absence of nonunitarity, quantum circuits composed of random unitary gates exhibit statistically symmetric spreading of local observables such as occupation in either direction~\cite{nahum2018operator,bertini2020scrambling,fisher2023random}. Over time, information initially localized in excitations disperses nearly uniformly across the system. Here, we demonstrate that this symmetry can be fundamentally broken by introducing feedback-based non-unitary operations. The resulting dynamics, emerging from the interplay between circuit randomness and feedback, give rise to a robust directional flow of information. Notably, this directionality arises from nonunitarity implemented at the circuit level via mid-circuit measurements and conditional operations. We refer to this framework as feedback-directed quantum dynamics.


To describe the effect of feedback-based control, we first introduce the projector acting on qubit $x$: $P_{x}[m]=\ket{m}\bra{m}$. If the outcome is $m$ for the $x$-th qubit, we apply a conditional operation $V_{x}[m]$. The associated Kraus operator is:
\begin{equation}
K_{x}[m]=V_{x}[m]P_{x}[m].
\end{equation}
In the quantum trajectory picture, the density matrix evolves as follows:
\begin{equation}
\rho\rightarrow \frac{K_{x}[m]\rho K_{x}^{\dagger }[m]}{p[m]}
\end{equation}
with $p[m]={\rm Tr}[P_{x}[m]\rho P_{x}[m]]$. This formalism makes the role of feedback explicit: once the system collapses onto a branch labeled by $m$, the deterministic control action $V_{x}[m]$ is applied. In the absence of feedback, $V_{x}[m]=I$, so the resulting dynamics reduce to random projections without directed influence. To incorporate a directional effect arising from feedback, $V_{x}[m]$ or $K_{x}[m]$ can be chosen asymmetrically across sites. Below, we describe two broad approaches for effecting direction-asymmetric feedback-based control:

\begin{itemize}
\item {\bf Position-dependent feedback:} Spatial asymmetry can be implemented by assigning feedback strengths that depend on the qubit position {$x$}:
\begin{equation}\label{fed1}
K_{x}[m]=f(x)V_{x}[m]P_{x}[m],
\end{equation} 
where $f(x)$ encodes position-dependent feedback strength and $V_x[m]$ is the conditional operation. Thus, the measurement record itself acts as a control signal, dynamically selecting the arrow of motion. 

\item{\bf {Feedback-conditioned operator:}}
Alternatively, asymmetry can be embedded directly into the feedback-conditioned operation itself:
\begin{equation}\label{fed2}
K_{x}[m]=a[m]O_{x,x+1}P_{x}[m],
\end{equation} 
where $O_{x,x+1}$ is a two-site operator, and the coefficient {$a[m]$} depends explicitly on the measurement outcome $m$. {That $O_{x,x+1}$ is explicitly asymmetric (linking $x$ to $x+1$ but not $x-1$)} also serves to asymmetrically steer transport and information flow correlated with the observed outcome.
\end{itemize}


\begin{figure*}
    \centering
    \includegraphics[width=.99\linewidth]{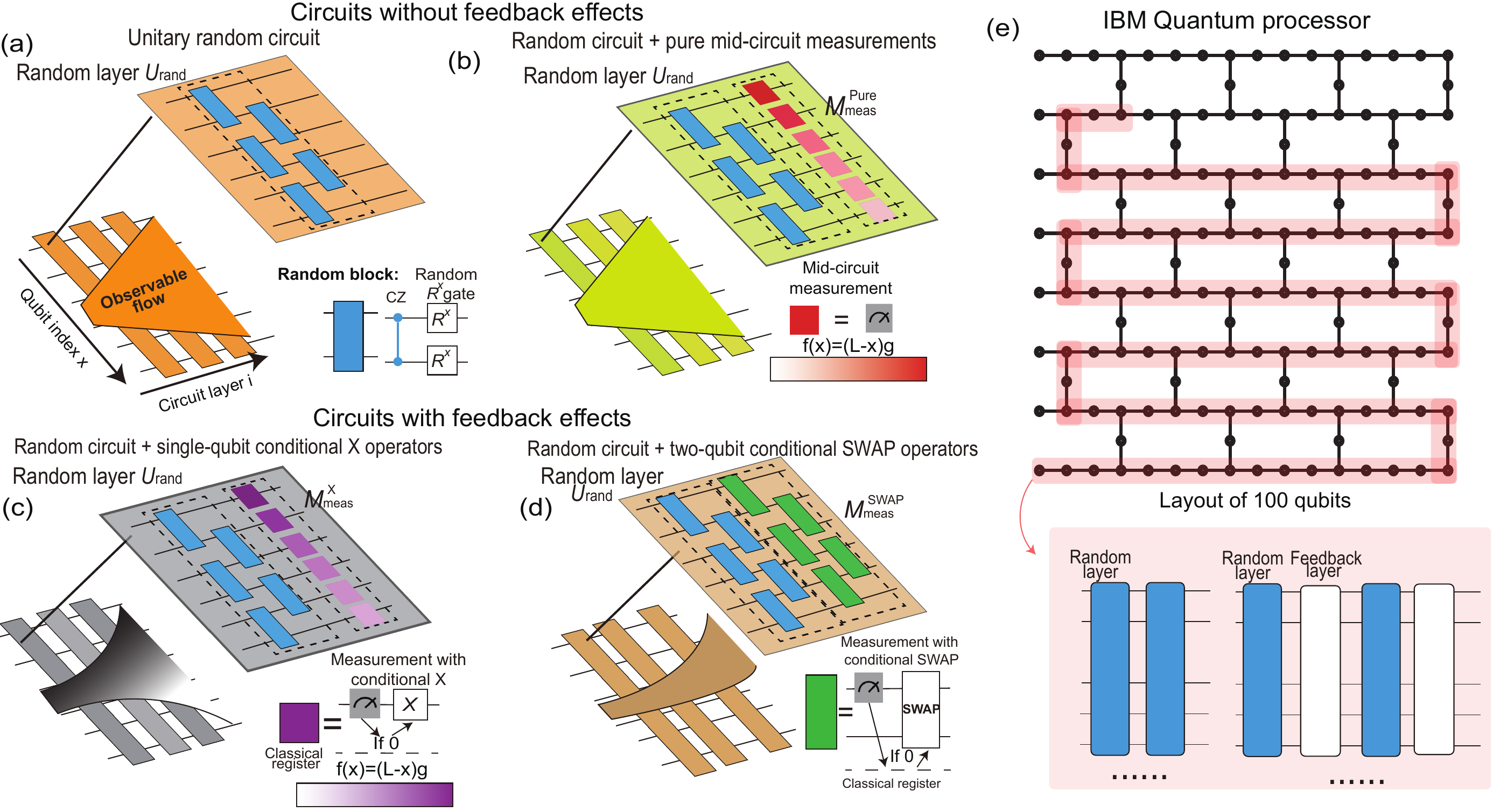}
    \caption{\textbf{Quantum circuit architectures for information scrambling and feedback-directed dynamics.}  
		(a) Baseline unitary random circuit  [Eq.~\ref{unitary}] built from local entangling blocks comprising random single-qubit rotations $R^x$ with CZ gates (light blue). The dynamics in this setup are purely unitary and exhibit no intrinsic directional bias, even as it spreads out due to scrambling. 
		(b) Random circuit with stochastic mid-circuit measurements (red) applied after the interlaced scrambling operations (blue) in each layer [Eq.~\ref{pure}].  Each qubit is measured independently with a spatially varying probability $f(x)=(L-x)g$ (red gradient), where $x$ labels the qubit position in an $L$-qubit system.  (c) Feedback circuit with single-qubit conditional-X gates (purple) [Eq.~\ref{x}]: in each layer, measurement outcomes are stored in a classical register and used to apply an outcome-conditioned X gate with probability $f(x)$ (purple gradient). (d) Feedback circuit with two-qubit conditional-SWAP operations: each green block, applied with a fixed probability $p^{\rm SWAP}$ [Eq.~\ref{swap}], represents a mid-circuit measurement followed by a SWAP between the measured qubit and its right neighbor if the outcome is $0$. 
   The funnel-shaped flows schematically illustrate the representative evolution of local observables in each setup. Only the feedback-based circuits shown in (c) and (d) exhibit asymmetric flows driven by position-dependent loss or effective drift. (e) Illustrative large-scale implementation on an IBM Quantum processor. The upper panel shows, for illustration, the selected 100-qubit chain on the IBM Quantum device, while the lower panel schematically depicts the corresponding circuit layout.}
    \label{fig:circuits}
\end{figure*}

\rz{Measurement-conditioned operations have been implemented previously for tasks such as state preparation and teleportation \cite{iqbal2024topological,google2023measurement}. Here, we employ measurement-conditioned operations as the central dynamical ingredient, thereby enabling a broader level of control over the flow of quantum information. In this work, we demonstrate these two feedback-based control approaches (Eqs.~\ref{fed1} and \ref{fed2}) by progressing through the four circuit setups shown in Figs.~\ref{fig:circuits}(a)–(d).
Each panel shows one iteration of the circuit configuration, where the input state is scrambled by random unitaries (blue) before undergoing various measurement schemes (red, purple, or green). }
 
Fig.~\ref{fig:circuits}(a) depicts a purely unitary circuit composed of random layers $U_{\rm rand}$ (blue), which serve to scramble the initial state only.
In Fig.~\ref{fig:circuits}(b), standard non-unitary mid-circuit measurements (red) are inserted after every random unitary layer (blue) with a qubit-dependent probability $f(x)$.
Next, Figs.~\ref{fig:circuits}(c) and (d) show the implementation of the two types of feedback described by Eqs.~\ref{fed1} and \ref{fed2}.
In Fig.~\ref{fig:circuits}(c), corresponding to Eq.~\ref{fed1}, each measurement block (purple) contains a conditional single-qubit $X$ operation. Spatial feedback asymmetry is introduced through the spatially varying measurement rate $f(x)$.
By contrast, in Fig.~\ref{fig:circuits}(d), corresponding to Eq.~\ref{fed2}, the feedback asymmetry is encoded directly in the feedback-controlled operation block (green), which contains a conditional two-qubit SWAP gate. No spatial inhomogeneity in the measurement rate 
is needed. Below, we further elaborate on each of these circuit configurations.

\subsubsection{Unitary random circuit without measurements and feedback}

As the most basic example, we consider the random quantum-circuit ansatz in Fig.~\ref{fig:circuits}(a), which serves to effect information scrambling \cite{nahum2017quantum,nahum2018operator,von2018operator}. Its architecture consists of repeated layers of unitary gates denoted as $U_{\rm rand}$, such that the overall evolution is given by
\begin{equation}\label{unitary}
U^{\rm unitary}(i)=[U_{\rm rand}]^{i}.
\end{equation}
Each layer (orange rectangle) comprises alternating even--odd arrangements (blue boxes) of controlled-$Z$ (CZ) gates and random single-qubit rotations about the $X$ axis, expressed as
\begin{equation}\label{rand}
U_{\rm rand}=\prod_{j\in \mathrm{even}}{CZ}_{j, j+1}R^X_{j}R^X_{j+1}\prod_{j\in \mathrm{odd}}{CZ}_{j, j+1}R^X_{j}R^X_{j+1},
\end{equation}
where $R^X_j = e^{-i\theta_j X/2}$ denotes a random rotation about the $X$ axis on qubit $j$. The Pauli operators are defined as $X=\ket{0}\bra{1}+\ket{1}\bra{0}$, $Y=-i\ket{0}\bra{1}+i\ket{1}\bra{0}$, and $Z=\ket{0}\bra{0}-\ket{1}\bra{1}$. In the following, we build upon this random-circuit ansatz in three increasingly sophisticated ways.

\subsubsection{Random circuit with pure mid-circuit measurements}

\rz{To generate non-unitary, measurement-induced dynamics, the most direct approach is to stochastically apply mid-circuit measurements (red boxes in Fig.~\ref{fig:circuits}(b)) after each random unitary layer (blue) at every iteration $i$. One layer of evolution is then described by the quantum channel
\begin{equation}\label{pure}
\rho \;\mapsto\; M_{\rm meas}^{\rm Pure}(\rho_{\rm scr}),
\qquad
\rho_{\rm scr}=U_{\rm rand}\rho U_{\rm rand}^{\dagger},
\end{equation}
where $\rho_{\rm scr}$ is the scrambled density matrix. Here, $M_{\rm meas}^{\rm Pure}
=
\prod_{x=0}^{L-1} M_{x}^{\rm Pure}$,
with the local single-qubit measurement channel at site $x$ defined by
\begin{equation}\label{eq:Mpure_local}
M_{x}^{\rm Pure}(\rho)
=
f(x)\sum_{m=0,1} P_x[m]\,\rho\,P_x[m]
+
\bigl(1-f(x)\bigr)\rho.
\end{equation}
Thus, with probability $f(x)$, qubit $x$ is projectively measured in the computational basis, while with probability $1-f(x)$ it is left unmeasured.}

{To set the stage for direction-dependent feedback later on, we allow the measurement rate $f(x)$ to be qubit-dependent. For concreteness, we fix it as the linear function}
\begin{equation}
f(x)=(L-x)g, \label{fi}
\end{equation}
throughout this work, where  {$x$} labels the qubit position,  {$g<1/L$} is a constant proportionality factor, and $L$ is the total length of the qubit chain, such that $f(x)$ vanishes linearly as $x$ approaches $L$. Specifically, at site $x$, the measurement channel acts on the density matrix as 
\begin{equation}\label{eq:meas_channel_single}
\rho \rightarrow  f(x)\left(P_{x}[0] {\rho_\text{scr}} P_{x}[0]+P_{x}[1] {\rho_\text{scr}} P_{x}[1]\right)+(1-f(x)) {\rho_\text{scr}}.
\end{equation}
Importantly, the operation in Eq.~\ref{eq:meas_channel_single} introduces no intrinsic directional selection {since the measurement rate $f(x)$ is only position-dependent, not direction-specific}. To see this, we examine how the local density expectation $\langle n_x\rangle = \text{Tr}\left[n_x\rho\right]$, where $n_{x}=\ket{0_{x}}\bra{0_{x}}$ is updated after one iteration. From Eq.~\ref{eq:meas_channel_single}, we see that 
\begin{align} \langle n_{x} \rangle &\rightarrow f(x)\text{Tr}[n_x\rho_\text{scr}] +(1-f(x))\text{Tr}[n_x\rho_\text{scr}] =\langle n_{x} \rangle_\text{scr}, \end{align} 
i.e. the updated density $\langle n_x\rangle$ follows that of the scrambled state, unaffected by $f(x)$. 
As a result, the dynamics generated by Eq.~\ref{pure} {must remain spatially symmetric (yellowish green triangle).}

\subsubsection{Random circuit with conditional single-qubit X operators}

\rz{To implement asymmetrically directed feedback, we now attach conditional operations to the mid-circuit measurements. We first consider the single-qubit feedback protocol of Eq.~\ref{fed1}, illustrated by the purple boxes in Fig.~\ref{fig:circuits}(c). In this protocol, if qubit $j$ is measured and the outcome is $0$, a conditional Pauli-$X$ gate is applied to flip the qubit. The corresponding outcome-$0$ Kraus operator is
\begin{equation}\label{eq:Kreset}
K^{\rm Reset}_{j}[0]
=
\sqrt{f(j)}\,X_{j}P_j[0],
\end{equation}
where $f(j)$ [Eq.~\ref{fi}] is the probability that qubit $j$ is measured. Since the full measurement-feedback step is a quantum channel rather than a single operator, one layer of evolution is written as
\begin{equation}\label{x}
\rho \;\mapsto\; M_{\rm meas}^{X}(\rho_{\rm scr}),
\qquad
\rho_{\rm scr}=U_{\rm rand}\rho U_{\rm rand}^{\dagger},
\end{equation}
where $M_{\rm meas}^{X}
=
\prod_{j=0}^{L-1} M_{j}^{X},$
with the local single-qubit measurement-feedback channel
\begin{equation}\label{eq:Mx_local}
\begin{aligned}
M_{j}^{X}(\rho)
={}&
f(j)\!\left[
X_{j}P_j[0]\,\rho\,P_j[0]X_{j}
+
P_j[1]\,\rho\,P_j[1]
\right]
\\
&+
\bigl(1-f(j)\bigr)\rho .
\end{aligned}
\end{equation}
Thus, after the local scrambling step $U_{\rm rand}$, the feedback channel stochastically redirects measured $\ket{0}$ outcomes into $\ket{1}$, while measured $\ket{1}$ outcomes remain unchanged and unmeasured qubits are left untouched.}

In terms of the local density $n_j$, this measurement process can be effectively modeled as a position-dependent loss term with a spatially asymmetric profile, with the following local-density update rule at site $j$:
\rz{\begin{equation}
\begin{aligned}
&\langle n_j\rangle \rightarrow
{\rm Tr}\Bigl[n_j\Bigl(
f(j)\,X_jP_j[0]\rho_{\rm scr}P_j[0]X_j \\
&+ f(j)\,P_j[1]\rho_{\rm scr}P_j[1]
+ (1-f(j))\,\rho_{\rm scr}
\Bigr)\Bigr]\\
&= (1-f(j))\langle n_j\rangle_{\rm scr}.
\end{aligned}
\end{equation}}
{Even if the scrambled density profile $\langle n_{x} \rangle_\text{scr}$ were spatially symmetric, the non-uniform measurement loss $f(x)$ would attenuate it unevenly, such that the updated density profile $\langle n_{x} \rangle$ would acquire directional bias, as depicted by the unevenly colored grey funnel-like shape. } 
In a nutshell, this measurement-conditioned $X$ feedback converts outcomes into an engineered, position-dependent channel, realizing controllable directed transport that is robust on large-scale noisy hardware, as we demonstrate later.

\subsubsection{Random circuit with conditional 2-qubit SWAP operators}

To realize asymmetrically directed feedback, we next consider a more sophisticated protocol based on two-qubit conditional-SWAP operations, as described by Eq.~\ref{fed2} and illustrated in Fig.~\ref{fig:circuits}(d). In this architecture, each green block denotes a measurement-conditioned operation applied with fixed probability $p^{\rm SWAP}$. Unlike the position-dependent measurement profile $f(x)$ used in the previous setups, here it is sufficient to use a spatially uniform measurement probability $p^{\rm SWAP}$. The outcome-$0$ Kraus operator acting on bond $(x,x+1)$ is
\begin{equation}\label{eq:Kswap}
K^{\rm SWAP}_{x}[0]
=
\sqrt{p^{\rm SWAP}}\,
{\rm SWAP}_{x,x+1}P_x[0].
\end{equation}
Since the full measurement-feedback step is a quantum channel rather than a single operator, one layer of evolution is written as
\begin{equation}\label{swap}
\rho \;\mapsto\; M_{\rm meas}^{\rm SWAP}(\rho_{\rm scr}),
\qquad
\rho_{\rm scr}=U_{\rm rand}\rho U_{\rm rand}^{\dagger},
\end{equation}
where the full measurement-feedback layer is given by the brickwork composition
\begin{equation}\label{eq:Mswap_def}
M_{\rm meas}^{\rm SWAP}
=
\left(\prod_{x\in{\rm even}} M_x^{\rm SWAP}\right)
\left(\prod_{x\in{\rm odd}} M_x^{\rm SWAP}\right),
\end{equation}
with the local bond channel
\begin{equation}\label{eq:Mswap_local}
\begin{aligned}
M_x^{\rm SWAP}(\rho)
={}&
p^{\rm SWAP}\![
{\rm SWAP}_{x,x+1}P_x[0]\,\rho\,P_x[0]{\rm SWAP}_{x,x+1}
\\
&+
P_x[1]\,\rho\,P_x[1]
]
+
\bigl(1-p^{\rm SWAP}\bigr)\rho .
\end{aligned}
\end{equation}
Here, with probability $p^{\rm SWAP}$, qubit $x$ is measured; if the outcome is $0$, a SWAP is applied between qubits $x$ and $x+1$, whereas if the outcome is $1$, no SWAP is applied. With the remaining probability $1-p^{\rm SWAP}$, the bond is left unchanged.

This conditional feedback operation effectively transfers the population in the state from {qubit} \( x \) to qubit \( x+1 \), thereby inducing a directed $x\rightarrow x+1$ flow of the \(|0\rangle\) component across the chain with likelihood proportional to the measurement probability $ {p^{\rm SWAP}}$. 
At the level of local densities, the exact update induced by a single conditional-SWAP channel on bond $(x,x+1)$ is (see Methods.~\ref{sec2m})
\begin{equation}
  \langle n_{x+1}\rangle\rightarrow \langle n_{x+1} \rangle +{p^{\rm SWAP}}[\langle n_{x} \rangle - \langle n_{x+1}n_{x}\rangle],  
\end{equation}
which indicates a forward hopping rate $p^{\rm SWAP}$ under $(\langle n_{x} \rangle - \langle n_{x+1}n_{x}\rangle)>0$.

Note that the origin of non-reciprocity in our Fig.~\ref{fig:circuits}(c) and (d) setups is fundamentally different from that underlying the conventional non-Hermitian skin effect (NHSE) \cite{yao2018edge,kunst2018biorthogonal,liang2022dynamic,yokomizo2019non}, which is also marked by non-unitary directional pumping. In the NHSE, directionality is typically encoded at the microscopic level through intrinsically asymmetric couplings or non-reciprocal hopping amplitudes \cite{okuma2020topological}. By contrast, our circuits require no built-in asymmetric couplers and no explicitly non-reciprocal unitary dynamics. Instead, non-reciprocity is generated operationally: measurement-conditioned feedback promotes stochastic measurement outcomes into outcome-dependent control actions that implement state-selective population transfer. The resulting asymmetry is therefore engineered at the level of quantum trajectories. Although the conditional operations are strictly local, the repeated coupling between the quantum state and its classical record can accumulate into genuinely collective behavior—for example, an effective global drift field (see Eq.~\ref{veffm} in Methods~\ref{sec:veff_swap}).

A key feature of our circuit architectures is their scalability. As illustrated in Fig.~\ref{fig:circuits} (e), these designs can be mapped naturally onto the connectivity of current IBM Quantum processors and extended to system sizes of up to $100$ qubits, and the robust feedback-directed dynamics will be demonstrated below. At present, the main obstacle to pushing these protocols to substantially greater depths is hardware latency: since mid-circuit readout and the subsequent feedforward operations introduce additional decoherence. As these readout and feedforward primitives improve in available quantum hardware, the same designs can be executed at significantly greater depths. A quantitative, calibration-based estimate of the attainable depth under current IBM  {quantum} hardware parameters is provided in Supplementary Note 6.

\subsection{ {Asymmetric dynamics from measurement-based feedback }}

 {Previously, we introduced circuit architectures [Figs.~\ref{fig:circuits}(c) and (d)] designed to effect feedback-directed dynamics. In this section, we provide initial digital quantum processor simulations that clearly illustrate that} the observed directional flow arises specifically from measurement-conditioned feedback operations, rather than from the measurement process.
Here we first present small-scale simulations with relatively weak noise and demonstrate their qualitative agreement with the dynamics from an effective stochastic model; larger-scale implementations that challenge the limitations of current quantum hardware will be presented afterwards in Section.~\ref{seclarge}.

\begin{figure*}
    \centering
       \includegraphics[width=0.99\linewidth]{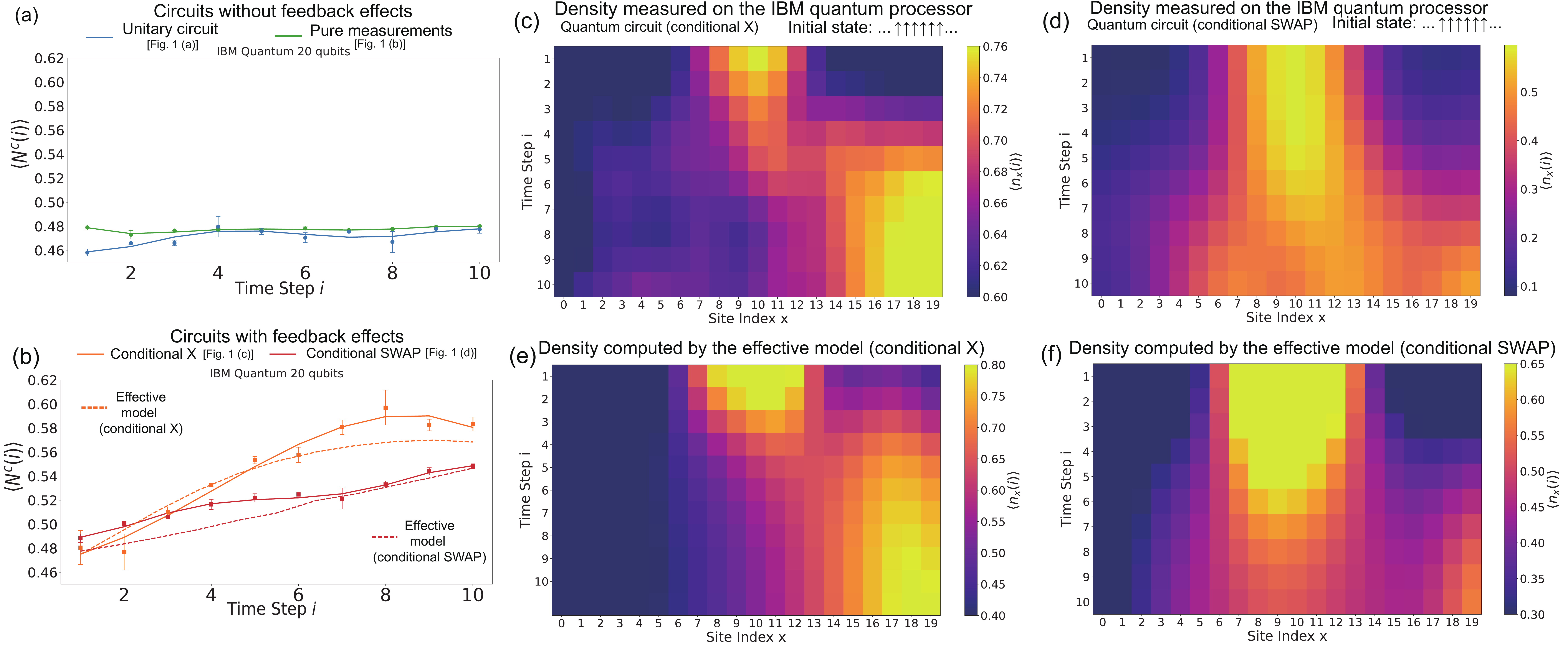}
\caption{\textbf{Demonstration of feedback-induced dynamics across the circuit architectures of Fig.~\ref{fig:circuits}.}
(a) Evolution of the center-of-mass observable [Eq.~\ref{den}] for chaotic unitary circuits (blue, Eq.~\ref{unitary}; Fig.~\ref{fig:circuits}(a)) and for circuits with position-dependent mid-circuit measurements only (green, Eq.~\ref{pure}; Fig.~\ref{fig:circuits}(b)). In both cases, the dynamics remain nearly symmetric, and no directed transport is observed.
(b) Evolution of the same observable for circuits with conditional $X$ feedback (yellow, Eq.~\ref{x}; Fig.~\ref{fig:circuits}(c)) and conditional SWAP feedback (red, Eq.~\ref{swap}; Fig.~\ref{fig:circuits}(d)). Conditional $X$ feedback produces a clear asymmetric flow, while the conditional SWAP protocol induces a weaker directional response. Dashed curves in (a) and (b) show results from stochastic models [Eq.~\ref{markov}; Methods~\ref{class}].
(c),(e) Space--time density profiles $\langle n_x(i)\rangle$ [Eq.~\ref{dent}] for the conditional-$X$ setup: IBM Quantum data in (c) and the corresponding stochastic simulation results [Eq.~\ref{markov}] in (e).
(d),(f) Space--time density profiles $\langle n_x(i)\rangle$ for the conditional-SWAP setup: IBM Quantum data in (d) and the corresponding results from the stochastic model [Eq.~\ref{markov}] in (f).
For all panels, the initial state consists of six initially occupied qubits localized near the center of a chain of length $L=20$. We use a measurement gradient $g=0.01$ and, for the SWAP protocol, $p^{\rm SWAP}=0.3$. Error bars in (a) and (b) denote the standard deviation over 10 random circuit realizations, and the solid curves are fitted to the measured data. All empirical data were obtained on the IBM Quantum processor \texttt{marrakesh}.
}
    \label{densitydy2}
\end{figure*}

To quantitatively characterize the asymmetric flow induced by feedback, we define the state density at qubit $x$ after the $i$-th layer iteration as
\begin{equation}\label{dent}
\langle {n}_x(i)\rangle=\text{Tr}[n_{x}\rho_{i}],
\end{equation}
as well as the center of mass of the evolved state
\begin{equation}\label{den}
\langle N^c(i)\rangle
=
\frac{\sum_x x\, {\rm Tr}\!\left[n_x{\rho}_{i}\right]}
{\sum_x {\rm Tr}\!\left[\,n_x{\rho}_{i}\right]}
=
\frac{\sum_x x\,\langle n_x(i)\rangle}
{\sum_x \langle n_x(i)\rangle}.
\end{equation}
Here, $\rho_{i}$ is the density matrix evolved after $i$ layers, $n_{x}=\ket{0_{x}}\bra{0_{x}}$ denotes the number-density operator at site $x$. Since the application of random $R^{x}$ rotations breaks particle-number conservation (i.e., magnetization conservation before the Jordan-Wigner transform), the total population $\sum_{x}\langle {n}_x(i)\rangle$ is no longer conserved. To account for this and ensure a meaningful interpretation of the density profile, we therefore include the normalization factor $\sum_{x}\langle {n}_x(i)\rangle$ in Eq.~\ref{den}.

\begin{figure*}
    \centering
    \includegraphics[width=0.98\linewidth]{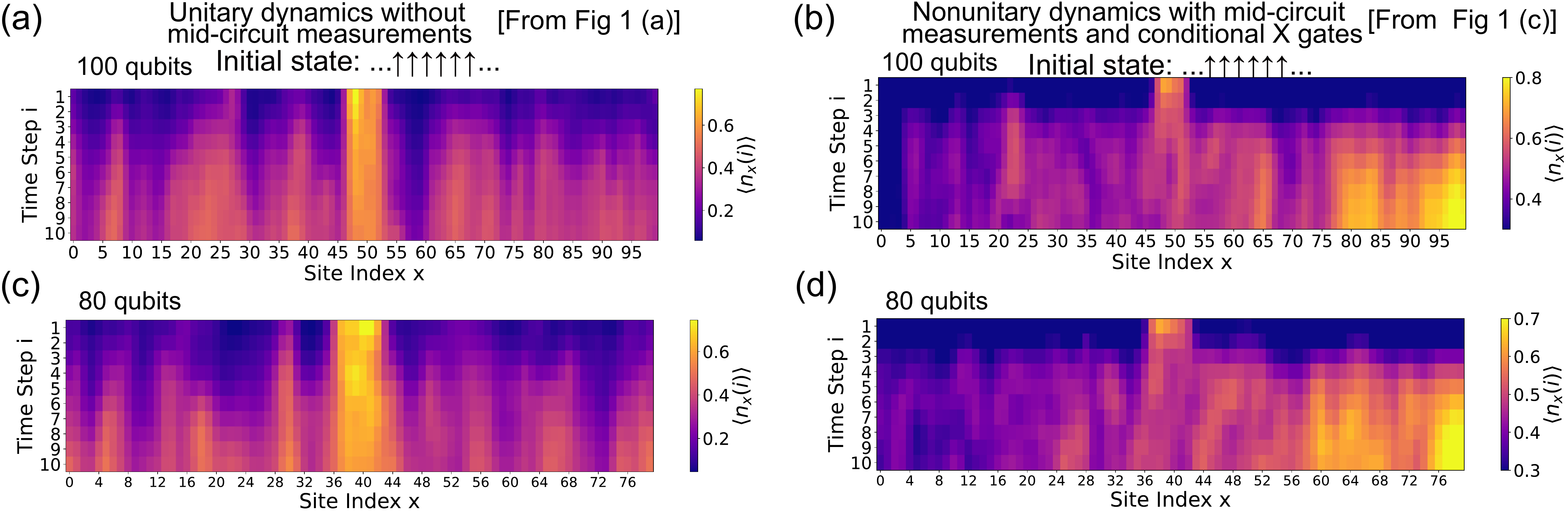}
    \caption{{Feedback-directed dynamics realized on large-scale quantum hardware with 80 to 100 qubits.} We present the local $n_{x}$ occupation over time steps $i$. The upper and lower panels present results for $100$- and $80$-qubit quantum simulations, respectively. In each case, the left panels correspond to chaotic unitary circuits [Eq.~\ref{unitary}] constructed from CZ and random $R^{x}$ gates [Eq.~\ref{rand}], and the right panels correspond to circuits with additional position-dependent mid-circuit measurement rate $f(x)=(L-x)g$ and conditional $X$ gates [Eq.~\ref{x}], as illustrated in Fig.~\ref{fig:circuits} (c). In the absence of mid-circuit measurements, the occupation profile remains symmetric, with no clear accumulation near the system boundaries. Conversely, when mid-circuit measurements with conditional-
X gates are introduced, an asymmetric flow emerges, leading to a visible accumulation of occupation. We set $g=0.004$ and $g=0.0015$ [Eq.~\ref{fi}] for $80$- and $100$-qubit simulations. The initial state in all cases consists of six spin-up qubits initialized at the center.
}
    \label{fig:large}
\end{figure*}

The measured dynamically evolved $\langle {N}^{c}(i)\rangle$ and $\langle {n}_x(i)\rangle$ on IBM Quantum hardware are presented in Fig.~\ref{densitydy2}(a,b) and (c-f) respectively.  In all cases, we consider a chain of length $L=20$ initialized with six occupied qubits localized near the center, and we track the evolution over circuit layers $i$. We first examine circuits with purely mid-circuit measurements [Eq.~\ref{pure}] (green curves) and those exhibiting fully unitary chaotic dynamics [Eq.~\ref{unitary}] (blue curves). In both cases [Fig.~\ref{densitydy2}(a)], the dynamics remain symmetric and exhibit no significant directional flow. By contrast, in Fig.~\ref{densitydy2}(b), the yellow curve corresponding to circuits with conditional $X$ gates exhibits an evident directional effect. A similar, though slightly weaker, feedback-directed motion is also observed in the SWAP-based circuits (red curves).

\subsubsection{Effective
 stochastic model for feedback-directed dynamics}

To further understand the measurement feedback dynamics, we provide a complementary approximate description of it from a stochastic perspective. In our measurement-based circuits, the dynamics between measurement events are largely governed by the unitary scrambling layers. Mid-circuit measurements primarily sample and update these local observables, so their backaction can be treated as a stochastic update of configuration probabilities. At the level of local observables, such as populations, the evolved states can be well approximated by an incoherent mixture in the computational basis. As such, the interplay of unitary scrambling and measurements can be captured quantitatively by an effective classical stochastic description (see Methods). Here, we model this effective process as a discrete-time stochastic process acting on bitstrings
$ b=(b_1,b_2,\dots,b_L)\in{0,1}^L $, leading to a time-dependent probability distribution $P(b)$. Each iteration layer implements a Markov update,
\begin{equation}\label{markov}
	 P_{i+1}(b') = \sum_{b} \mathcal{M}(b'\!\leftarrow\! b)\, P_i(b).  
\end{equation}
Here, $\mathcal{M}$ is the transition kernel for one full layer, including scrambling and measurements (see Methods~\ref{class} for more details).

In our {circuit constructions [Figs.~\ref{fig:circuits}(c-d)], each layer consists of two components: the scrambling component followed by the measurement feedback component.} The first component redistributes, i.e., scrambles information locally by updating neighboring pairs of bits according to transition probabilities obtained from the action of the underlying two-qubit quantum layer $(R_{x}\otimes R_{x})CZ$. The second component is a measurement–feedback map: each classical bit is sampled (``measured''), and an outcome-conditioned update is applied to emulate the corresponding conditional operation in the quantum circuit. All results of effective models shown in the main text refer to simulations of the discrete-time Markov process in Eq.~\ref{markov}, averaged over many stochastic trajectories and random realizations (see Methods.~\ref{class}).

At least for shallow circuits (fewer than 10 layers), this classical stochastic description (dashed curves) provides an effective representation of the corresponding measured quantum dynamics (solid curves), as demonstrated in the $\langle N^C(i)\rangle$ plots of Figs.~\ref{densitydy2}(a)–(b). Moreover, the measured  {qubit}-resolved density profiles $\langle {n}_x(i)\rangle$ shown in Figs.~\ref{densitydy2}(c)–(d) also show quantitative agreement with the classical stochastic predictions [Figs.~\ref{densitydy2}(e)–(f)] obtained from trajectory averaging (see Methods.~\ref{class}).
As expected, the two approaches result in distinct asymmetric dynamics: In the conditional flip approach, the effective position-dependent loss $\gamma(x)$ yields a relatively smooth monotonically increasing density profile; while in the conditional SWAP approach, the effective drift results in rather obvious directional skewing motion of the density profile.

\rz{Within the classical description of Eq.~\ref{markov}, taking the continuum limit ${n}_x(i)\rightarrow n(x,t)$, the two feedback architectures reduce to distinct mechanisms as derived in Methods~\ref{class} and \ref{sec:veff_swap}. Measurement-conditioned flips selectively remove the 0 population, which appears as an effective, position-dependent loss rate  {$\gamma(x)$ proportional to the local measurement strength $f(x)$}. Moreover, in the conditional-SWAP architecture, the outcome-conditioned action generates a systematic right-moving bias, which manifests as an emergent drift field with velocity $v\propto p^{\rm swap}$.}

\subsection{Quantum simulations of feedback effects in large-scale systems}\label{seclarge}

Previous results in Fig.~\ref{densitydy2} show that the dynamics under feedback-controlled $X$ gates remain qualitatively robust against both gate imperfections and measurement noise in the hardware simulation. Moreover, the additional randomness introduced by hardware noise can be partially averaged out by the intrinsic randomness of the circuit itself, as discussed in the Supplementary Information. Leveraging  {on} this robustness, we extend our simulations beyond small system sizes to explore the scalability of the architecture depicted in Fig.~\ref{fig:circuits}(c).

Below we demonstrate that feedback-directed dynamics can be realized reliably at large scales on current noisy quantum hardware, extending up to an unprecedented 100 qubits, as shown in Fig.~\ref{fig:large}. In Fig.~\ref{fig:large}(a), we present, {for comparison}, the purely unitary evolution for systems with 80 and 100 qubits. In the absence of mid-circuit measurements and feedback, the local density profiles remain nearly symmetric throughout the evolution, in analogy with the behavior observed at smaller system sizes and consistent with the lack of any intrinsic directional bias. When spatially structured mid-circuit measurements are combined with conditional feedback operations as described in Fig.~\ref{fig:circuits}(c), a pronounced asymmetry {and subsequent edge accumulation develop in the density profiles, as shown in Fig.~\ref{fig:large}(b). Remarkably, this feedback-induced directional flow persists even at these very large scales, where the increased system size naturally amplifies the impact of noise. The continued emergence of a clear asymmetric response at 80–100 qubits therefore demonstrates that the measurement feedback mechanism remains effective.

\rz{Although imperfect gate control introduces effectively random microscopic errors, their net effect is largely washed out by the intrinsic randomness of the circuit ensemble, so that the emergent directed response remains robustly observable even at large system sizes. We further emphasize that the large-scale results are averaged over multiple random realizations (10 in this work), which, together with the intrinsic randomness within each circuit, provide sufficient stochastic averaging to reduce the influence of individual gate imperfections. This is supported by Fig.~\ref{fig:large}(b), where the 80- and 100-qubit data exhibit similarly robust asymmetric trends despite the increased circuit complexity. The persistence of this behavior highlights the scalability of our approach for controlling non-unitary dynamics on quantum hardware.}

\subsubsection{Quantitative characterization of the extent of dynamical feedback}

We next quantitatively characterize the measured dynamical evolution in the 100-qubit chain subject to conditional spin flips. To isolate initial boundary effects, we consider the net center-of-mass shift at iteration (time) $i$
\begin{equation}\label{den2}
\delta {N}^{c}(i)=L(\langle {N}^{c}(i)\rangle-\langle {N}^{c}(1)\rangle),
\end{equation} 
where $L$ is the system size.
We simulate the conditional-$X$ feedback architecture [Eq.~\ref{x}] on 100-qubit chains with two distinct initial configurations: one with six spin-up qubits (i.e. occupied fermions) initialized at the center of the chain [Fig.~\ref{fig:ace}(a)], and another with 6 spin-up qubits localized near the right edge [Fig.~\ref{fig:ace}(b)]. As shown in Fig.~\ref{fig:ace}(a), the center-initialized state (yellow curve) exhibits a pronounced directional displacement driven by the conditional $X$ gates, whereas the corresponding unitary circuit and the circuit with measurements but no feedback (blue and green curves) {exhibit} negligible direction propagation. 
When the initial state is prepared near the right edge in a region with a low measurement rate [Fig.~\ref{fig:ace}(b)], both the feedback-driven circuit and the measurement-only circuit exhibit largely localized, frozen dynamics. In this regime, repeated measurements tend to project the system onto boundary-localized states and suppress effective scrambling.
In particular, this setup demonstrates that the feedback-based circuit does not simply act as an effective loss mechanism; rather, it relies on the position-dependent selection from the measurement outcomes across quantum trajectories, and can be blocked by a spatial boundary.

\begin{figure*}
    \centering
    \includegraphics[width=0.85\linewidth]{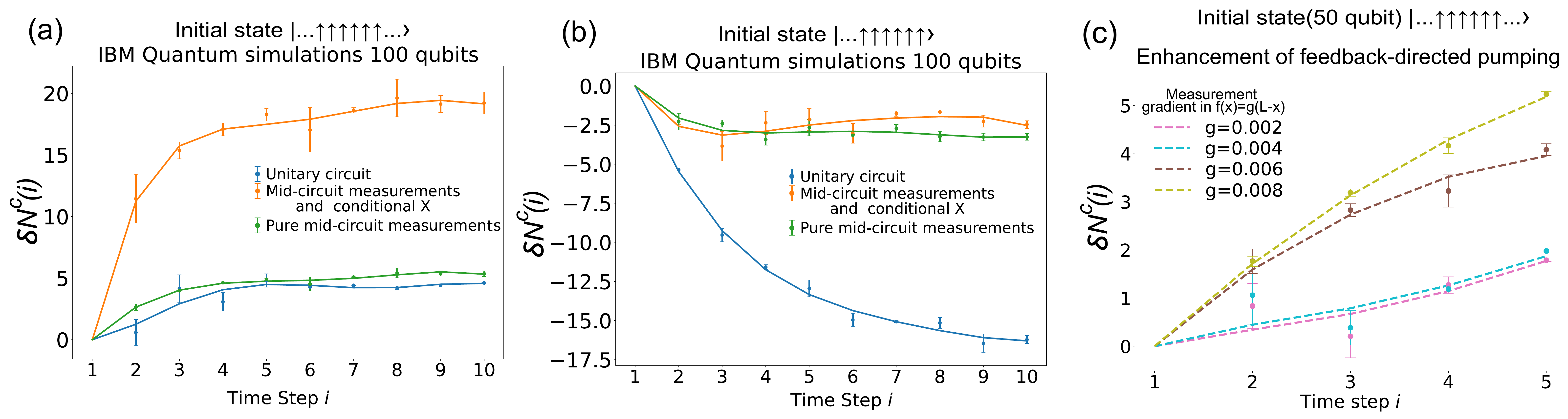}
		\caption{\textbf{Quantifying feedback-directed pumping in large-scale quantum hardware simulations.}
		We quantify feedback-directed dynamics through the net center-of-mass shift in qubit position defined in Eq.~\ref{den2}. Panels (a) and (b) show simulations initialized with six spin-up qubits prepared at the center and near the right edge, respectively. Blue curves represent results from unitary circuits composed only of CZ and $R^{x}$ gates [Eq.~\ref{unitary}], with a single layer described by Eq.~\ref{rand}. Yellow curves correspond to simulations on circuits with position-dependent mid-circuit measurements and conditional $X$ operations [Eq.~\ref{x}] (Fig.~\ref{fig:circuits}(c)), and green curves represent simulations on random circuits with only mid-circuit measurements [Eq.~\ref{pure}] (Fig.~\ref{fig:circuits}(b)). In (a), feedback induces a pronounced enhancement of rightward density flow (yellow curve). In (b), the dynamics are restricted by the right boundary. In the region ($x\rightarrow L$ for $f(x)=g(L-x)$) with low measurement rates, feedback has a negligible effect. For all results, we set the gradient of the measurement probability to $g=0.15/L$, with $L$ denoting the total number of qubits. In (c), we implement the controlled-$X$ circuit (50 qubits) [Eq.~\ref{x}] and measure the net center-of-mass shift for different circuit depths $i$ and gradients of the measurement probability $g$ [Eq.~\ref{fi}]. The circuit size is $L=50$. For slightly deeper circuits, the measured net center-of-mass exhibits enhanced pumping under increasing gradients $g$. 
		}
    \label{fig:ace}
\end{figure*}

\rz{Moreover, in our setup, the feedback-induced asymmetric flow is not an uncontrolled chaotic response; rather, it is a tunable dynamical effect set by the strength of the measurement-and-feedback channel. We therefore quantify its dependence on the measurement probability $f(x)=(L-x)g$ controlled by the gradient parameter $g$. Due to hardware constraints that limit the number of mid-circuit measurements, we focus on a representative system size of $L=50$ qubits. In Fig.~\ref{fig:ace}(c), we present the evolution of the center-of-mass shift [Eq.~\ref{den2}] for increasing values of $g$. A clear monotonic trend emerges: as $g$ increases, the density profile is displaced more strongly. In particular, even a modest increase of 
 $g$ from $0.004$ to $0.006$ produces a visible enhancement of the response, demonstrating that the mid-circuit feedback acts as an active control setup in circuits.}

\subsubsection{Measurements of robust transport signatures}

The above results demonstrate that mid-circuit measurements combined with conditional operations can generate a robust directional flow that remains observable up to system sizes of 100 qubits. For larger systems, however, the dynamics naturally spread over an extended bulk region, so the feedback-induced bias can become less apparent in site-averaged observables, and can be partially masked by bulk measurement imperfections. To sensitively quantify directional transport in this large-scale regime and mitigate the impact of bulk measurement imperfections, we introduce a polarization-like diagnostic that measures the boundary-to-boundary population imbalance,
\begin{equation}\label{cur}
\langle J^{c}(i)\rangle
=
\frac{{\rm Tr}\!\left[\,(n_{L-1}-n_{0}){\rho}_{i}\right]}
{\sum_x {\rm Tr}\!\left[\,n_x{\rho}_{i}\right]}
=
\frac{\langle n_{L-1}(i)\rangle-\langle n_{0}(i)\rangle}
{\sum_x \langle n_x(i)\rangle}.
\end{equation}
which directly compares the occupations at the two ends of the chain {$x=0$ and $x=L-1$}. For symmetric dynamics, $\langle J^{c}(i) \rangle$ remains close to zero. By contrast, in feedback-directed circuits\CH{,} this quantity exhibits a pronounced response, even when the local bulk density profiles appear nearly symmetric. Thus, the effective polarization provides a robust and size-insensitive signature of directed flow, enabling reliable detection of feedback-induced transport in the large-scale regime.

As shown in Fig.~\ref{fig:density}(a), the measured effective polarization extracted from fully unitary circuits [Eq.~\ref{unitary}] remain close to zero for all three system sizes, consistent with the absence of any intrinsic directional bias. The small residual deviations arise from finite sampling over random circuit realizations and do not affect the qualitative conclusion. By contrast, Fig.~\ref{fig:density}(b) presents the results for the conditional-$X$ feedback circuit [Eq.~\ref{x}] using the size-scaled measurement gradient $g=0.18/L$. Across all system sizes, the measured effective polarization displays a consistently increasing 
trend, providing clear evidence that feedback induces robust directional flow. Notably, when the dynamics are quantified using the effective polarization rather than local density profiles alone, the directional response becomes substantially more pronounced, with $\langle J^{c}(i) \rangle$ reaching values as large as $\approx 0.6$. This enhancement reflects the fact that this diagnostic directly aggregates the net boundary-to-boundary redistribution of population, yielding a clear and size-robust signature of feedback-directed transport. Importantly, by compressing the spatial response into a single imbalance measure, it helps ensure that the directional signal is not suppressed by noise or bulk averaging at large scales.

\begin{figure*}
    \centering
    \includegraphics[width=0.69\linewidth]{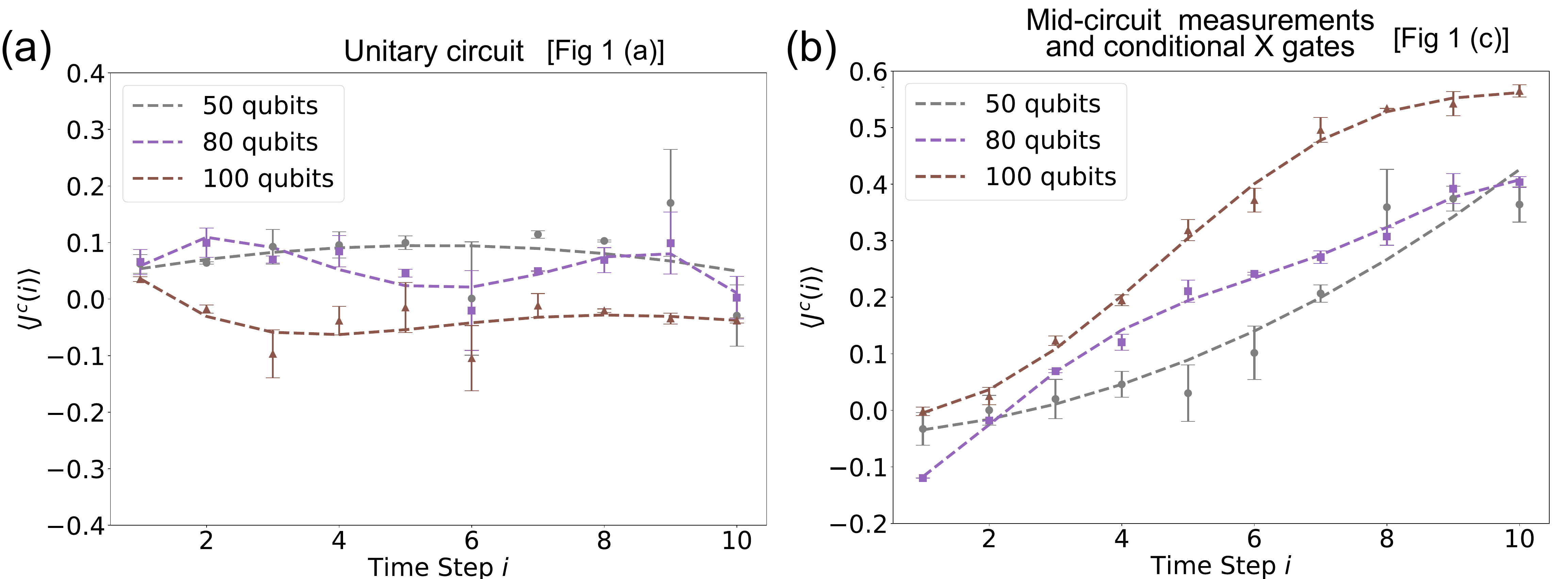}
    \caption{{Compact end-to-end polarization as a scalable diagnostic of feedback-directed transport.} Panels (a) and (b) show the end-to-end polarization defined in Eq.~\ref{cur}, evaluated from quantum hardware simulations of purely unitary circuits [Eq.~\ref{unitary}, panel (a)] and circuits with conditional $X$ feedback [Eq.~\ref{x}, panel (b)].  This observable provides a minimal global diagnostic that compresses the large-scale dynamics into a single quantity while remaining sensitive to feedback-induced directional transport.  Results are shown for system sizes of 50, 80, and 100 qubits, shown as red, purple, and brown curves, respectively. In panel (b), we use the measurement gradient $g=0.18/L$. The comparison between panels (a) and (b) shows that, whereas the purely unitary dynamics produce little net end-to-end polarization, conditional $X$ feedback generates a clear imbalance across the chain. In all cases, the initial state consists of six spin-up qubits prepared at the center of the chain. Error bars indicate the mean deviation over independent 10 random circuit realizations, and solid curves represent fits to these measured data.
		}
    \label{fig:density}
\end{figure*}

\section{Discussion}

Our study introduces and simulates a new mechanism for directing quantum dynamics in chaotic many-body systems using spatially arranged mid-circuit measurements.  A key result of our work is the {demonstration of directional non-unitary dynamics at the unprecedentedly large scale of up to 100 qubits on digital quantum hardware. }
The observed asymmetric pumping and occupation shifts  {persist robustly without much degradation even at 100 qubits}, demonstrating that feedback control can be implemented in a scalable manner on existing quantum hardware. From a conceptual standpoint, our study highlights that feedback from mid-circuit measurements can transform monitored dynamics into programmable control resources, bridging measurement-induced physics and engineered non-unitary behavior.

Importantly, our findings are significant not only for understanding measurement-induced dynamics but also for establishing a scalable platform for simulating non-Hermitian physics using digital quantum devices. We show that the resource requirements of our setup scale linearly with system size, making execution on hundreds of qubits feasible. Moreover, the present performance ceiling is set primarily by mid-circuit readout latency and by readout and feedback errors. As these hardware primitives improve—most notably through faster, higher-fidelity mid-circuit measurement and feedforward—the same architecture can be pushed to substantially greater depths, enabling access to longer-time non-unitary many-body dynamics. Such long-time, programmable non-unitary evolution are particularly challenging to realize and diagnose in other leading platforms, including cold-atom systems. A quantitative, calibration-based estimate of the attainable depth on IBM hardware is provided in Supplementary Note 6.

The programmability of our approach naturally extends beyond directed transport. By tailoring the temporal structure of the measurement and feedback layers, the same platform can be used to engineer and probe other nonequilibrium phenomena, including dynamical phase transitions and measurement-induced entanglement transitions~\cite{vijay2012stabilizing,baumer2024quantum,baumer2024efficient,skinner2019measurement,nahum2021measurement,google2023measurement}. Moreover, implementing feedback directly at the circuit level provides a concrete route to active, measurement-based control of many-body quantum evolution, enabling information flow and entanglement structure to be designed and controlled in space and time through programmable measurement patterns and real-time conditional operations.

\section*{Methods}

  \subsection{Mitigating errors in dynamical circuits}

In our setup, mid-circuit measurements and conditional feedforward introduce additional error channels beyond those of purely unitary circuits. In particular, imperfect readout and finite-latency feedforward can accumulate over layers. We therefore optimize the measurement arrangement used in our simulations. Concretely, prior to hardware execution, we generate a pool of candidate sampling {schemes} 
that specify which qubits are measured at each circuit layer. We then benchmark these patterns using calibration information and retain only those that minimize the readout error. This pre-selection step improves the robustness of simulations.

More broadly, current hardware imposes a practical constraint on the total number of mid-circuit measurements that can be executed. This constraint limits the maximum measurement-probability gradient that can be implemented for the feedback circuit [Eq.~\ref{x}]. As a result, the quantum simulations reported in the main text focus on moderate gradient values and relatively shallow depths, where feedback-induced directionality remains resolvable. A quantitative estimate is provided in Supplementary Note 6.

For the SWAP-based implementation, each SWAP gate is compiled into the native gate set using the standard decomposition $\operatorname{SWAP}_{i, i+1}=[\operatorname{CZ}_{i,i+1}\sqrt{X}_{i}\sqrt{X}_{i+1}]^{3}$. This can lead to temporal sparsity, since SWAP operations are applied only to a subset of neighboring pairs in each layer. To mitigate this effect, we insert identity operations, $I=X^{2}$, on qubits where SWAP gates are not applied, so as to reduce coherence loss. Concrete estimates based on our particular hardware implementations are also discussed in Supplementary Note 6.

\subsection{Density dynamics of quantum feedback circuits with scrambling and conditional SWAP gates}\label{sec2m}

In the SWAP-based feedback circuit [Eq.~\ref{swap}], the measurement-conditioned operation on the bond $(x,x+1)$ is described by the local quantum channel
\begin{equation}
\mathcal E_{x,x+1}(\rho)
=
K_0\rho K_0^\dagger
+
K_1\rho K_1^\dagger
+
K_{\rm id}\rho K_{\rm id}^\dagger,
\end{equation}
with Kraus operators $K_0=\sqrt{p^{\rm SWAP}}\,{\rm SWAP}_{x,x+1}P_x[0],
K_1=\sqrt{p^{\rm SWAP}}\,P_x[1],
K_{\rm id}=\sqrt{1-p^{\rm SWAP}}\,I.$
Here $P_x[0]=\ket{0}\bra{0}$ and $P_x[1]=\ket{1}\bra{1}$ project onto the measured outcomes at qubit $x$. Thus, with probability $p^{\rm SWAP}$, qubit $x$ is measured; if the outcome is $0$, a SWAP is applied between $x$ and $x+1$, whereas if the outcome is $1$, no SWAP is applied. With the remaining probability $(1-p^{\rm SWAP})$, the bond is left unchanged.

\rz{Accordingly, we have
\begin{equation}
\begin{aligned}
&\langle n_{x+1}\rangle'
=
{\rm Tr}\!\left[n_{x+1}\,\mathcal{E}_{x,x+1}(\rho)\right] \\
&=
{\rm Tr}\!\left[n_{x+1}K_0\rho K_0^\dagger\right]
+
{\rm Tr}\!\left[n_{x+1}K_1\rho K_1^\dagger\right]
+
{\rm Tr}\!\left[n_{x+1}K_{\rm id}\rho K_{\rm id}^\dagger\right] \\
&=
p^{\rm SWAP}\,{\rm Tr}\!\left[n_{x+1}\,{\rm SWAP}_{x,x+1}P_x[0]\rho P_x[0]{\rm SWAP}_{x,x+1}\right] \\
&\quad
+p^{\rm SWAP}\,{\rm Tr}\!\left[n_{x+1}P_x[1]\rho P_x[1]\right]
+
(1-p^{\rm SWAP})\,{\rm Tr}[n_{x+1}\rho].
\end{aligned}
\label{eq:nx1_kraus_expand}
\end{equation}
Using the trace and the identities, ${\rm SWAP}_{x,x+1}^\dagger n_{x+1}{\rm SWAP}_{x,x+1}=n_x,
P_x[0]=n_x,
P_x[1]=I-n_x $, 
we can express the first term as
\begin{equation}
\begin{aligned}
&{\rm Tr}\!\left[n_{x+1}\,{\rm SWAP}_{x,x+1}P_x[0]\rho P_x[0]{\rm SWAP}_{x,x+1}\right]\\
&=
{\rm Tr}\!\left[n_x P_x[0]\rho P_x[0]\right] \\
&=
\langle n_x\rangle,
\end{aligned}
\label{eq:first_term_nx1}
\end{equation}
where we used $P_x[0]=n_x$ and $n_x^2=n_x$. For the second term, we have
\begin{equation}
\begin{aligned}
{\rm Tr}\!\left[n_{x+1}P_x[1]\rho P_x[1]\right]=
\langle n_{x+1}(1-n_x)\rangle.
\end{aligned}
\label{eq:second_term_nx1}
\end{equation}
Substituting Eqs.~\eqref{eq:first_term_nx1} and \eqref{eq:second_term_nx1} into Eq.~\eqref{eq:nx1_kraus_expand}, we obtain
\begin{equation}
\begin{aligned}
\langle n_{x+1}\rangle'=
\langle n_{x+1}\rangle
+
p^{\rm SWAP}\Bigl[\langle n_x\rangle-\langle n_xn_{x+1}\rangle\Bigr].
\end{aligned}
\label{eq:nx1_update_detailed}
\end{equation}}

For the density on site $x$, we have
\begin{equation}
\begin{aligned}
&\langle n_x\rangle'
=
{\rm Tr}\!\left[n_x\,\mathcal E_{x,x+1}(\rho)\right] \\
&=
{\rm Tr}\!\left[n_xK_0\rho K_0^\dagger\right]
+
{\rm Tr}\!\left[n_xK_1\rho K_1^\dagger\right]
+
{\rm Tr}\!\left[n_xK_{\rm id}\rho K_{\rm id}^\dagger\right] \\
&=
p^{\rm SWAP}\,{\rm Tr}\!\left[n_x\,{\rm SWAP}_{x,x+1}P_x[0]\rho P_x[0]{\rm SWAP}_{x,x+1}\right] \\
&\quad
+p^{\rm SWAP}\,{\rm Tr}\!\left[n_xP_x[1]\rho P_x[1]\right]
+
(1-p^{\rm SWAP})\,{\rm Tr}[n_x\rho].
\end{aligned}
\label{eq:nx_kraus_expand}
\end{equation}
Using ${\rm SWAP}_{x,x+1}^\dagger n_x{\rm SWAP}_{x,x+1}=n_{x+1}$,
we then express the first term as
\begin{equation}
\begin{aligned}
&{\rm Tr}\!\left[n_x\,{\rm SWAP}_{x,x+1}P_x[0]\rho P_x[0]{\rm SWAP}_{x,x+1}\right]\\
&=
{\rm Tr}\!\left[n_xn_{x+1}n_x\rho\right] \\
&=
{\rm Tr}\!\left[n_xn_{x+1}\rho\right] \\
&=
\langle n_xn_{x+1}\rangle.
\end{aligned}
\label{eq:first_term_nx}
\end{equation}
For the second term, we obtain
\begin{equation}
\begin{aligned}
{\rm Tr}\!\left[n_xP_x[1]\rho P_x[1]\right]
=
{\rm Tr}\!\left[(n_xP_x[1])\rho P_x[1]\right]
=0.
\end{aligned}
\label{eq:second_term_nx}
\end{equation} Substituting Eqs.~\eqref{eq:first_term_nx} and \eqref{eq:second_term_nx} into Eq.~\eqref{eq:nx_kraus_expand}, we obtain
\begin{equation}
\begin{aligned}
\langle n_x\rangle'
&=
p^{\rm SWAP}\,\langle n_xn_{x+1}\rangle
+
(1-p^{\rm SWAP})\langle n_x\rangle \\
&=
\langle n_x\rangle
-
p^{\rm SWAP}\Bigl[\langle n_x\rangle-\langle n_xn_{x+1}\rangle\Bigr].
\end{aligned}
\label{eq:nx_update_detailed}
\end{equation}
The conditional SWAP therefore induces a net forward transfer from $x$ to $x+1$ whenever site $x$ is occupied and site $x+1$ is empty, producing an effective right-moving bias.

\subsection{Classical stochastic models for local scrambling and controlled operations}\label{class}

Here, we introduce a stochastic model that captures the essential features of the quantum simulation studied in the main text. Note that this discrete Markov description should be distinguished from the continuum equations introduced later in Methods~\ref{medpde}: the latter provide only a phenomenological interpretation of dynamics.

We represent the system by an ensemble of classical bitstrings of length $L$,
\begin{equation}
b=(b_1,b_2,\dots,b_L)\in\{0,1\}^L,
\end{equation}
where each bit $b_x$ records the local computational-basis state at site $x$. At the effective circuit layer $i$, the state of the classical model is specified by a probability distribution $P_i(b)$. For any observable $O(b)$ depending only on the classical configuration, its expectation value is
\begin{equation}
\langle O\rangle_i=\sum_b O(b)\,P_i(b).
\end{equation}
In practice, these expectation values are estimated by averaging over many sampled stochastic trajectories, and we use $200$ in this work.

To model the quantum circuits shown in  Figs.~\ref{fig:circuits}(c,d), each classical layer of our stochastic model consists of two stages: a scrambling stage followed by a measurement--feedback stage. Accordingly, one layer is represented by the Markov update
\begin{equation}\label{markovsupp}
P_{i+1}(b')=\sum_b \mathcal M(b'\!\leftarrow\! b)\,P_i(b),
\qquad
\mathcal M=\mathcal R\circ\mathcal S,
\end{equation}
where $\mathcal S$ denotes the scrambling kernel and $\mathcal R$ the feedback kernel.

\paragraph{Scrambling stage.}

The scrambling stage is designed to mimic the local redistribution of basis populations generated by the two-qubit quantum block $(R_x\otimes R_x)\,CZ$. For a bond $(x,x+1)$, with input configuration $(a,b)\in\{0,1\}^2$ and output configuration $(c,d)\in\{0,1\}^2$, we define the local transition matrix
\begin{equation}\label{eq:T_def}
T^{(x,x+1)}_{(a,b)\rightarrow(c,d)}
=
\big|\langle c,d|(R_x\otimes R_x)\,CZ|a,b\rangle\big|^2.
\end{equation}
Thus, $T^{(x,x+1)}$ is a stochastic two-site channel obtained directly from the computational-basis transition probabilities of the corresponding quantum gate block.

To build the full scrambling step on the chain, we apply these local channels in a brickwork pattern. We set $\mathcal B_{\rm odd}=\{(1,2),(3,4),\dots\}$ and $\mathcal B_{\rm even}=\{(2,3),(4,5),\dots\}$, which denote the odd and even bonds. For $\mu\in\{{\rm odd},{\rm even}\}$, we define
\begin{equation}
\mathcal S_\mu(b'\!\leftarrow\! b)
=
\prod_{(x,x+1)\in\mathcal B_\mu}
T^{(x,x+1)}_{(b_x,b_{x+1})\rightarrow(b'_x,b'_{x+1})},
\end{equation}
so that all bonds within the same sublayer are updated independently. The full scrambling kernel is then represented as the following ordered composition
\begin{equation}
\mathcal S=\mathcal S_{\rm even}\circ \mathcal S_{\rm odd}.
\end{equation}

\paragraph{Classical stochastic conditional-$X$ feedback [Eq.~\ref{x}].}

We first consider the classical counterpart of the conditional-$X$ feedback circuit. At each site $x$, the bit is reset to $1$ with probability $f(x)$:
\begin{equation}\label{eq:reset_rule}
b_x\mapsto
\begin{cases}
1, & \text{with probability } f(x),\\
b_x, & \text{with probability } 1-f(x).
\end{cases}
\end{equation}
The corresponding site-local kernel is
\begin{equation}
\mathcal R_x^{X}(b'_x\!\leftarrow\! b_x)
=
f(x)\,\delta_{b'_x,1}
+
\bigl[1-f(x)\bigr]\delta_{b'_x,b_x},
\end{equation}
and the full feedback kernel factorizes as
\begin{equation}
\mathcal R^{X}=\prod_x \mathcal R_x^{X}.
\end{equation}
In this work, for consistency, we take a linear spatial profile $f(x)=g(L-x)$.

\paragraph{Classical stochastic conditional-SWAP feedback [Eq.~\ref{swap}].}
We next consider the classical counterpart of the conditional-SWAP feedback circuit. On each bond $(x,x+1)$, the update take an exchange with probability $p^{\rm SWAP}$ conditioned on the left bit being $0$:
\begin{equation}\label{eq:swap_rule}
(b_x,b_{x+1})\!\!\mapsto\!\!
\begin{cases}
(b_{x+1},b_x), &\!\!\!\!\text{with probability } p^{\rm SWAP}\ \text{if}\ b_x=0,\\
(b_x,b_{x+1}), &\!\!\!\text{otherwise}.
\end{cases}
\end{equation}
These updates are applied in a brickwork pattern over non-overlapping bonds, consistent with Fig.~\ref{fig:circuits} (d), and the entire update process can be presented
\begin{equation}
\mathcal R^{\rm SWAP}
=
\mathcal R^{\rm SWAP}_{\rm even}\circ \mathcal R^{\rm SWAP}_{\rm odd}.
\end{equation}

In particular, the dashed curves in Fig.~\ref{densitydy2}(a,b) and the effective-model density profiles compared with the IBM data in Fig.~\ref{densitydy2}(c\text{--}f) are generated from repeated simulations of Eq.~\eqref{markov}. To reduce statistical fluctuations and isolate robust dynamical features, we average observables over $W$ independent stochastic realizations:
\begin{equation}
\overline{\langle O\rangle_i}
=
\frac{1}{W}\sum_{r=1}^{W}\langle O\rangle_i^{(r)}.
\end{equation}
In practice, we use $W=200$ for the classical stochastic simulations.

\subsection{Continuum limit version of the effective classical models}\label{medpde}
We further elaborate on the classical models introduced above by deriving their continuum descriptions.  In this subsection, we denote by $n_j^{(\ell)}\in[0,1]$ the probability that qubit $j$ is in state $0$ after the $\ell$-th circuit layer. We use $j$ for the discrete qubit index and $\ell$ for the discrete layer index, so as to avoid conflict with the notation used in the main text. To connect the discrete stochastic description of Eq.~\ref{markov} to a continuum picture, we then introduce continuum variables for this and the following sections
\begin{equation}
x = j\,\Delta x,\qquad t=\ell\,\Delta t,\qquad n_j^{\ell}\rightarrow n(x,t),
\end{equation}
where $\Delta x$ is the lattice spacing and $\Delta t$ is the time associated with one effective layer.

For the conditional-$X$ architecture, the exact update derived above implies that, at the level of one-point densities $n(x,t+\Delta t) \rightarrow \bigl[1-f(x)\bigr]\,n(x,t)$,
where $f(x)$ is the probability that qubit $x$ is measured and, conditioned on the outcome. Writing
\begin{equation}
f(x)=\gamma(x)\,\delta t,
\end{equation}
we have
\begin{equation}
\frac{n(x,t+\Delta t)-n(x,t)}{\Delta t}
=
-\gamma(x)\,n(x,t)+\mathcal O(\Delta t).
\end{equation}
Taking $\Delta t\to 0$ gives the continuum loss equation
\begin{equation}
\partial_t n(x,t)=-\gamma(x)\,n(x,t).
\label{eq:x_loss_continuum}
\end{equation}

The scrambling part of the circuit produces additional spatial mixing. We therefore add an effective diffusion term and write
\begin{equation}\label{pdeg}
\partial_t n(x,t)= -\gamma(x)\,n(x,t)+D[n].
\end{equation}
Equation~\eqref{pdeg} is a coarse-grained effective description rather than an exact microscopic identity. In the numerical reconstructions used for the conditional-$X$ data, we employ the minimal decay--diffusion ansatz $D[n]=D\,\partial_x^2 n,$
with $D$ fitted directly to the measured density profiles; see Supplementary Note~5 and Fig.~S3 for the fitted reconstruction.

\subsection{Effective velocity induced by conditional-SWAP feedback}\label{sec:veff_swap}

The conditional SWAP on bond $(j,j+1)$ transfers population from $j$ to $j+1$ with current
\begin{equation}
J_{j\to j+1}^{(\ell)}
=
p^{\rm SWAP}\,
 n_j^{(\ell)}\bigl(1-n_{j+1}^{(\ell)}\bigr).
\label{eq:bond_current_discrete}
\end{equation}
Likewise, the neighbouring bond $(j-1,j)$ contributes an incoming current
\begin{equation}
J_{j-1\to j}^{(\ell)}
=
p^{\rm SWAP}\,
 n_{j-1}^{(\ell)}\bigl(1-n_j^{(\ell)}\bigr).
\label{eq:bond_current_incoming}
\end{equation}
Therefore, at the level of a discrete continuity equation, the net update of the local density at site $j$ is
\begin{equation}
n_j^{(\ell+1)}-n_j^{(\ell)}
=
J_{j-1\to j}^{(\ell)}-J_{j\to j+1}^{(\ell)}.
\label{eq:discrete_continuity_exact}
\end{equation}
Substituting Eqs.~\eqref{eq:bond_current_discrete} and \eqref{eq:bond_current_incoming} gives
\begin{equation}
\begin{aligned}
n_j^{(\ell+1)}-n_j^{(\ell)}
&=
p^{\rm SWAP}
\Bigl[
 n_{j-1}^{(\ell)}(1-n_j^{(\ell)})
-
 n_j^{(\ell)}(1-n_{j+1}^{(\ell)})
\Bigr].
\end{aligned}
\label{eq:discrete_continuity_with_correlations}
\end{equation}

We here consider the reduction
\begin{equation}
n_{j-1}^{(\ell)}\bigl(1-n_j^{(\ell)}\bigr)\approx n_{j-1}^{(\ell)},
\qquad
n_j^{(\ell)}\bigl(1-n_{j+1}^{(\ell)}\bigr)\approx n_j^{(\ell)},
\end{equation}
so that
\begin{equation}
n_j^{(\ell+1)}-n_j^{(\ell)}
\approx
p^{\rm SWAP}\Bigl[n_{j-1}^{(\ell)}-n_j^{(\ell)}\Bigr].
\label{eq:swap_linearized_discrete}
\end{equation}

Under $n_j^{(\ell)}\rightarrow n(x,t)$ and $n_{j-1}^{(\ell)}\rightarrow n(x-\Delta x,t)$, Eq.~\eqref{eq:swap_linearized_discrete} can be interpreted as a forward-time discrete update,
\begin{equation}
n(x,t+\Delta t)-n(x,t)
\approx
p^{\rm SWAP}\Bigl[n(x-\Delta x,t)-n(x,t)\Bigr].
\label{eq:swap_ftbs}
\end{equation}
To get a continuum description,  we expand both sides in $\Delta t$ and $\Delta x$:
\begin{equation}
n(x,t+\Delta t)
=
n(x,t)+\Delta t\,\partial_t n(x,t)+\mathcal O(\Delta t^2),
\label{eq:taylor_time}
\end{equation}
and
\begin{equation}
n(x-\Delta x,t)
=
n(x,t)-\Delta x\,\partial_x n(x,t)
+\frac{\Delta x^2}{2}\,\partial_x^2 n(x,t)
+\mathcal O(\Delta x^3).
\label{eq:taylor_space}
\end{equation}
Substituting Eqs.~\eqref{eq:taylor_time} and \eqref{eq:taylor_space} into Eq.~\eqref{eq:swap_ftbs} gives
\begin{equation}
\begin{aligned}
\Delta t\,\partial_t n(x,t)
\approx
-p^{\rm SWAP}\Delta x\,\partial_x n(x,t)
\\+\frac{p^{\rm SWAP}\Delta x^2}{2}\,\partial_x^2 n(x,t)
+\mathcal O(\Delta x^3,\Delta t^2).    
\end{aligned}
\end{equation}
Dividing by $\Delta t$ then gives
\begin{equation}
\partial_t n(x,t)
\approx
-\frac{p^{\rm SWAP}\Delta x}{\Delta t}\,\partial_x n(x,t)
+\cdots.
\label{eq:swap_continuum_corrected}
\end{equation}
Therefore, at leading order, the conditional-SWAP update generates an effective drift velocity
\begin{equation}\label{veffm}
v_{\rm eff}
=
\frac{p^{\rm SWAP}\Delta x}{\Delta t}.
\end{equation}
More details are shown in Supplementary Note~5 and Fig.~S4.

\section*{Author contributions} R.~S. proposed the initial idea for this work and implemented simulations on the IBM Quantum devices.  C.~H.~L. supervised the project and provided some theoretical input. 
All authors contributed to the preparation of the manuscript. 

\section*{ Data Availability Statement}
All data from this work are available from the corresponding authors upon reasonable request.

\section*{ Code Availability Statement}
All codes used in this work are available from the corresponding authors upon request. 
\section*{ Competing Interests}

The authors declare no competing interests.

\section*{ACKNOWLEDGMENTS}
We acknowledge the use of IBM Quantum services for this work. The views expressed are those of the authors and do not reflect the official policy or position of IBM or the IBM Quantum team. All data and codes of this work are available from the corresponding authors upon reasonable request. This work is supported by the Singapore Ministry of Education Academic Research Fund Tier-II Grant (Award No. MOE-T$2$EP$50222$-$0003$) and Tier-I preparatory grant (WBS no. A-8002656-00-00).

\bibliography{reference}

\onecolumngrid
\flushbottom
\newpage
\appendix
\setcounter{equation}{0}
\setcounter{figure}{0}
\setcounter{table}{0}
\setcounter{section}{0}
\renewcommand{\theequation}{S\arabic{equation}}
\renewcommand{\thefigure}{S\arabic{figure}}
\renewcommand{\thesection}{S\arabic{section}}
\renewcommand{\thepage}{S\arabic{page}}

\newpage
\section*{Supplementary \CH{Information}}

\section*{Supplementary Note 1. Classical matrix product state simulations of circuits with mid-circuit measurements}

Here, we present our tensor-network method for simulating quantum circuits with mid-circuit measurements. We represent the many-qubit wavefunction as a matrix product state (MPS) and evolve it layer by layer. Each circuit layer is decomposed into two sequential parts,
\begin{equation}
U^{(\ell)} = U_{\rm measurement}^{(\ell)} U_{\rm random}^{(\ell)},
\end{equation}
where $U_{\rm random}^{(\ell)}$ denotes the random unitary scrambling layer at circuit step $\ell$, and $U_{\rm measurement}^{(\ell)}$ denotes the subsequent mid-circuit measurement layer. In practice, $U_{\rm random}^{(\ell)}$ is applied to the MPS by sequentially updating the corresponding one- and two-qubit gates, with standard bond-dimension truncation after each two-site update.

After the unitary step, we perform projective measurements in the computational ($Z$) basis on the selected qubits. For a measured qubit $i$, we define the projectors
\begin{equation}
P_i^{(0)}=\ket{0_i}\bra{0_i}=\frac{1+Z_i}{2},
\qquad
P_i^{(1)}=\ket{1_i}\bra{1_i}=\frac{1-Z_i}{2}.
\end{equation}
Given the evolved MPS $\ket{\psi}$, the probabilities for the two outcomes are
\begin{equation}
p_i^{(0)}=\bra{\psi}P_i^{(0)}\ket{\psi}
=\frac{1+\langle Z_i\rangle}{2},
\qquad
p_i^{(1)}=\bra{\psi}P_i^{(1)}\ket{\psi}
=\frac{1-\langle Z_i\rangle}{2},
\end{equation}
where $\langle Z_i\rangle=\bra{\psi}Z_i\ket{\psi}$ is obtained directly from the local MPS expectation value. We then sample the measurement outcome $s_i\in\{0,1\}$ according to the distribution $\{p_i^{(0)},p_i^{(1)}\}$ and update the state by the corresponding projection,
\begin{equation}
\ket{\psi}\;\longrightarrow\;
\frac{P_i^{(s_i)}\ket{\psi}}{\sqrt{p_i^{(s_i)}}}.
\end{equation}
The MPS is re-normalized before proceeding to the next measured site.

This trajectory-based procedure yields one stochastic realization of the monitored circuit. To obtain physical observables, we repeat the full evolution over many independent realizations of both the random unitary layers and the sampled measurement outcomes, and then average the resulting observables over the ensemble.

\section*{Supplementary Note 2. Threshold randomness strength for reliable simulation of feedback-driven scrambling}

In this note, we examine how much unitary randomness is needed for the hardware dynamics to faithfully reproduce the intended feedback-driven dynamics.

We consider the same {conditional-$X$ feedback protocol} used in the main text. Each circuit layer consists of two parts. First, a random scrambling unitary is applied,
\begin{equation}
U_{\rm rand}^{(i)}
=
U_{\rm CZ}\,
\prod_{j=1}^{L} R_j^x(\theta_{j,i}),
\label{eq:urand_supp}
\end{equation}
where $R_j^x(\theta)=e^{-i\theta X_j/2}$ is a single-qubit rotation about the $x$ axis on qubit $j$, and the angles $\theta_{j,i}$ are independently sampled for each qubit $j$ and layer $i$ from a uniform interval $\theta_{j,i}\in[0,\theta_{\max}].$
Here, $U_{\rm CZ}$ denotes the nearest-neighbour CZ entangling layer used in the scrambling block. After this scrambling step, we apply the position-dependent measurement-feedback operation of the conditional-$X$ circuit: qubit $x$ is measured with probability $f(x)=g(L-x)$,
and, conditioned on the measurement outcome, a local $X$ gate is applied as in the protocol discussed in the main text. The parameter $g$ therefore sets the spatial gradient of the measurement probability across the chain.

To assess the role of scrambling strength, we compare two choices of the rotation-angle interval, $\theta_{\max}=0.5$ and $\theta_{\max}=0.8$, and benchmark the measured IBM Quantum results against  numerical simulations (see Supplementary Note 1), as shown in Fig.~\ref{fig:random}. In both cases, the underlying evolution protocol is the same; only the randomness strength of the scrambling block is varied.

As shown in Fig.~\ref{fig:random}(a), when the rotation angles are restricted to the narrower interval $[0,0.5]$, the measured dynamics become unreliable: the hardware data display an apparent asymmetric pumping signal that is not reproduced by the noiseless simulation. This indicates that, in the weak-scrambling regime, the circuit does not mix local information strongly enough to wash out coherent hardware imperfections, so spurious device-dependent fluctuations can generate false directional signatures. By contrast, increasing the scrambling strength improves the agreement between experiment and ideal numerics. As shown in Fig.~\ref{fig:random}(b), enlarging the rotation range to $[0,0.8]$ produces much better quantitative consistency with the noiseless simulation. In this stronger-scrambling regime, the random unitary layer more effectively redistributes local information and suppresses the impact of systematic hardware imperfections on the measured large-scale dynamics. Based on this threshold behavior, we choose a broader interval, $[0,1.0]$, for the simulation in the main text. This ensures that the circuit operates in a sufficiently strong scrambling regime.

\begin{figure}[h]
    \centering
    \includegraphics[width=0.75\linewidth]{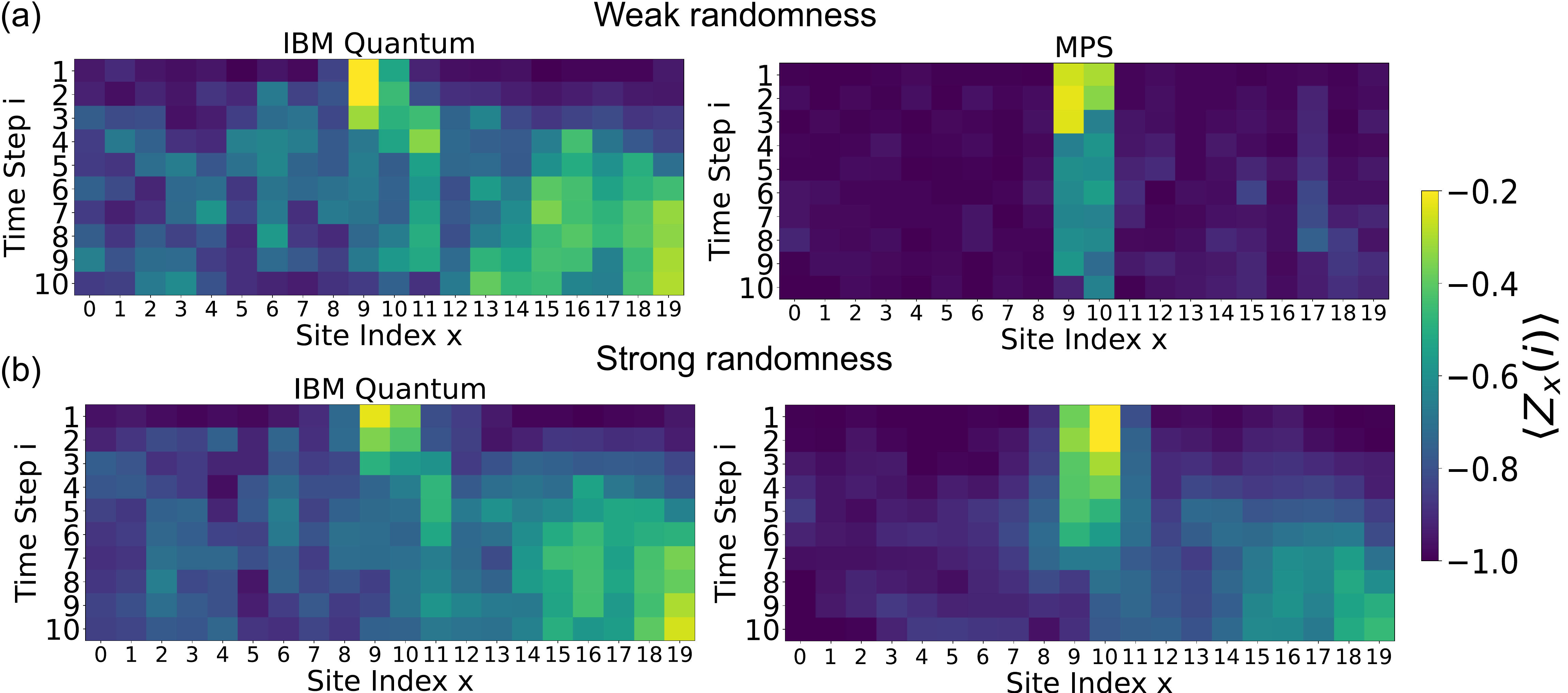}
    \caption{Threshold randomness strength test required for reliable implementation of the conditional-$X$ feedback protocol on noisy hardware. In each circuit layer, a random scrambling unitary $U_{\rm rand}^{(i)}=U_{\rm CZ}\prod_j R_j^x(\theta_{j,i})$ is applied first, where the angles $\theta_{j,i}$ are independently sampled from a uniform interval, followed by position-dependent mid-circuit measurement and conditional $X$ feedback with measurement profile $f(x)=g(L-x)$. Panels compare IBM Quantum results (left) with noiseless numerical simulations (right) for two different ranges of rotation angles: (a) $\theta_{j,i}\in[0,0.5]$ and (b) $\theta_{j,i}\in[0,0.8]$. In the weaker-scrambling regime of panel (a), the measured dynamics show spurious asymmetric behavior not present in the ideal simulation, indicating that hardware imperfections are insufficiently averaged out. In panel (b), the stronger scrambling suppresses these artifacts and restores much better agreement between hardware and noiseless results. For all panels, the measurement-probability gradient is set by $g=0.01$, corresponding to $f(x)=g(L-x)$ on a chain of length $L$. All data are under a single random realization.}
    \label{fig:random}
\end{figure}

\section*{Supplementary Note 4. Noise information of the IBM Quantum hardware}\label{ibmq}
\begin{figure}[h]
    \centering
    \includegraphics[width=0.9\linewidth]{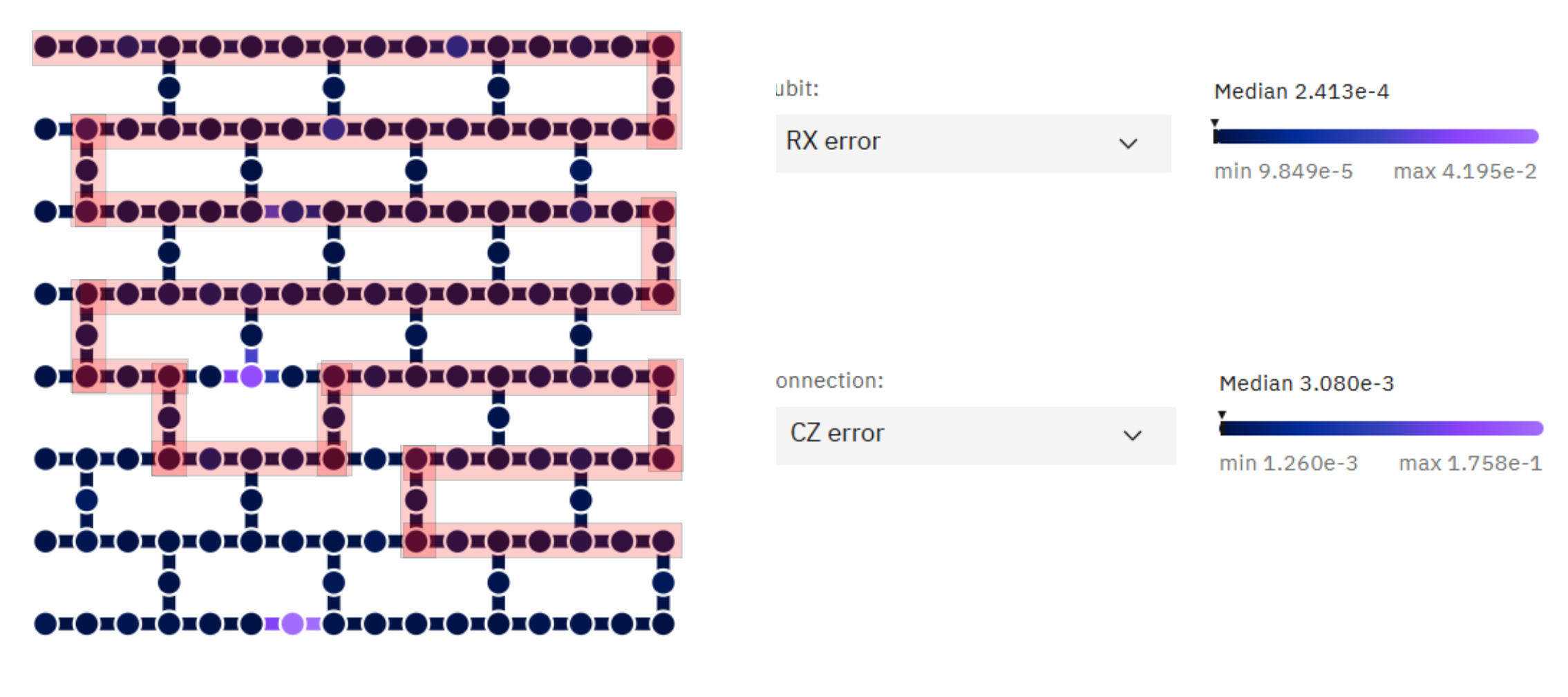}
    \caption{Error profile of the IBM Quantum device. The color bars on the right indicate the calibrated error rates for the native single-qubit $R^{X}$ gate and the two-qubit CZ gate. The 100-qubit chain in red identifies a low-noise region. }
    \label{fig:device}
\end{figure}

For the simulations in this work, we use the IBM Quantum device ``marrakesh", with the native $R^{X}$ and CZ gates. The error profile of this device is shown in Fig.~\ref{fig:device}. The qubit layout is automatically selected using the AI-assisted tool provided by the IBM Quantum platform to optimize circuit performance, which identifies a qubit chain with low noise (see the qubit chain in the red region).

\section*{Supplementary Note 5. Diffusion ansatz for effective stochastic models of measurement feedback processes}
\begin{figure}
	\centering
	\includegraphics[width=0.5\linewidth]{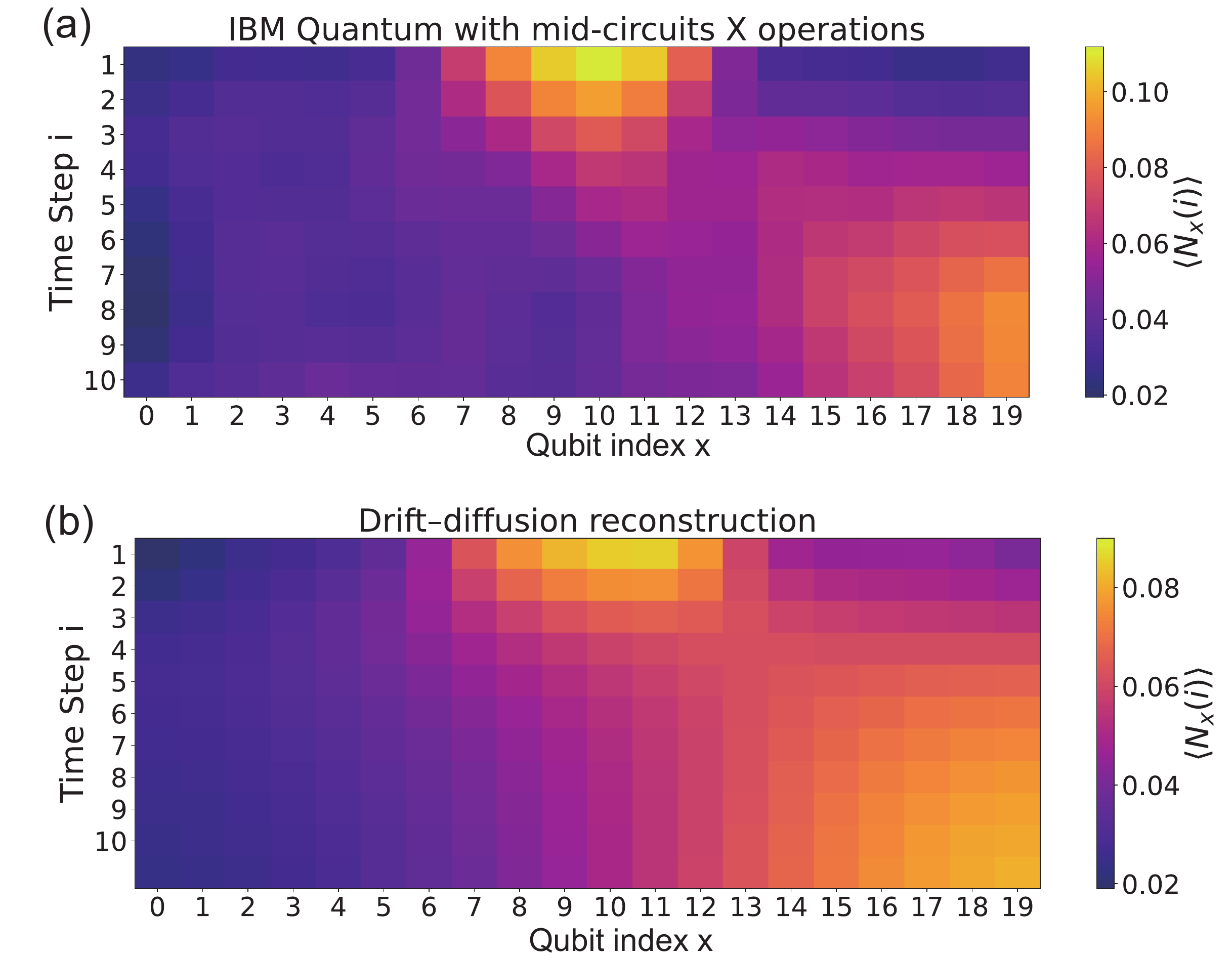}
	\caption{(a) Measured space–time density profile obtained from IBM Quantum hardware, showing the evolution of local occupations under feedback-directed dynamics. These data are adopted from Fig.2 (main text), and the density is normalized as. (b) Decay–diffusion reconstruction of the same evolution, obtained by fitting the measured data to the effective continuum model in Eq.~\ref{pde} with parameters $\gamma=0.26$ and $D=6.38$. The reconstructed profile captures the directional flow observed on the quantum processor.}
\label{fig:classicaldata}
\end{figure}

In the main text and Methods Sect.~D, we show that the dynamics generated by the conditional-$X$ feedback circuit can be described, at a coarse-grained level, by a decay--diffusion continuum model. The purpose of this section is to make explicit the form and interpretation of the diffusion contribution $D[N]$, which phenomenologically captures the spatial mixing generated by the scrambling unitaries. The resulting effective equation reads
\begin{equation}
\partial_t N(x,t)=-(1-x)\gamma\,N(x,t)+D[N],
\end{equation}
where $N(x,t)$ denotes the normalized local density profile. To obtain a minimal closed description, we approximate the diffusion functional by the standard Laplacian form
\begin{equation}\label{pde}
\partial_t N(x,t)=-(1-x)\gamma\,N(x,t)+D\,\partial_x^2 N(x,t).
\end{equation}
Here, the first term represents the position-dependent loss induced by the measurement-conditioned $X$ operations, while the second term models the effective spatial spreading caused by information scrambling in the random unitary layers. Note that Eq.~\eqref{pde} is not intended as an exact evolution equation, but rather as a phenomenological ansatz. The coefficients $\gamma$ and $D$ are extracted by fitting the measured density profiles.

Throughout this analysis, we introduce a dimensionless continuum coordinate
\begin{equation}
u=\frac{j}{L}\in[0,1],
\end{equation}
where $j=0,\dots,L-1$ labels the discrete lattice sites. The continuum density $N(u,t)$ is normalized as
\begin{equation}
\int_0^1 N(u,t)\,du=1,
\end{equation}
and is related to the discrete normalized site density $N_j(t)$ through
\begin{equation}
N_j(t)\approx \int_{(j-1)/L}^{j/L} N(u,t)\,du.
\end{equation}
As shown in Fig.~\ref{fig:classicaldata}, this minimal decay--diffusion model provides a quantitatively reasonable description of the observed asymmetric density evolution.

To describe the conditional-SWAP feedback circuit, a simple diffusion equation is no longer sufficient. We therefore introduce a higher-order (order-$k$) drift--diffusion expansion to capture the conditional-SWAP circuit. The effective evolution of the density is written as
\begin{equation}\label{pdedirft}
\partial_t N
=
-v\,\partial_x N
+\sum_{j=2}^{k} D_j\,\partial_x^{\,j} N,
\end{equation}
where $v$ is the emergent drift velocity generated by the measurement-conditioned SWAP rule, and the coefficients $D_j$ quantify higher-order spatial spreading arising from the interplay of scrambling and conditional SWAP. Here, we find that retaining terms up to order $k=4$ is sufficient to reproduce the measured density profile, as demonstrated by the results shown in Fig.~\ref{fig:classicaldata2}.

\begin{figure}
    \centering
    \includegraphics[width=0.5\linewidth]{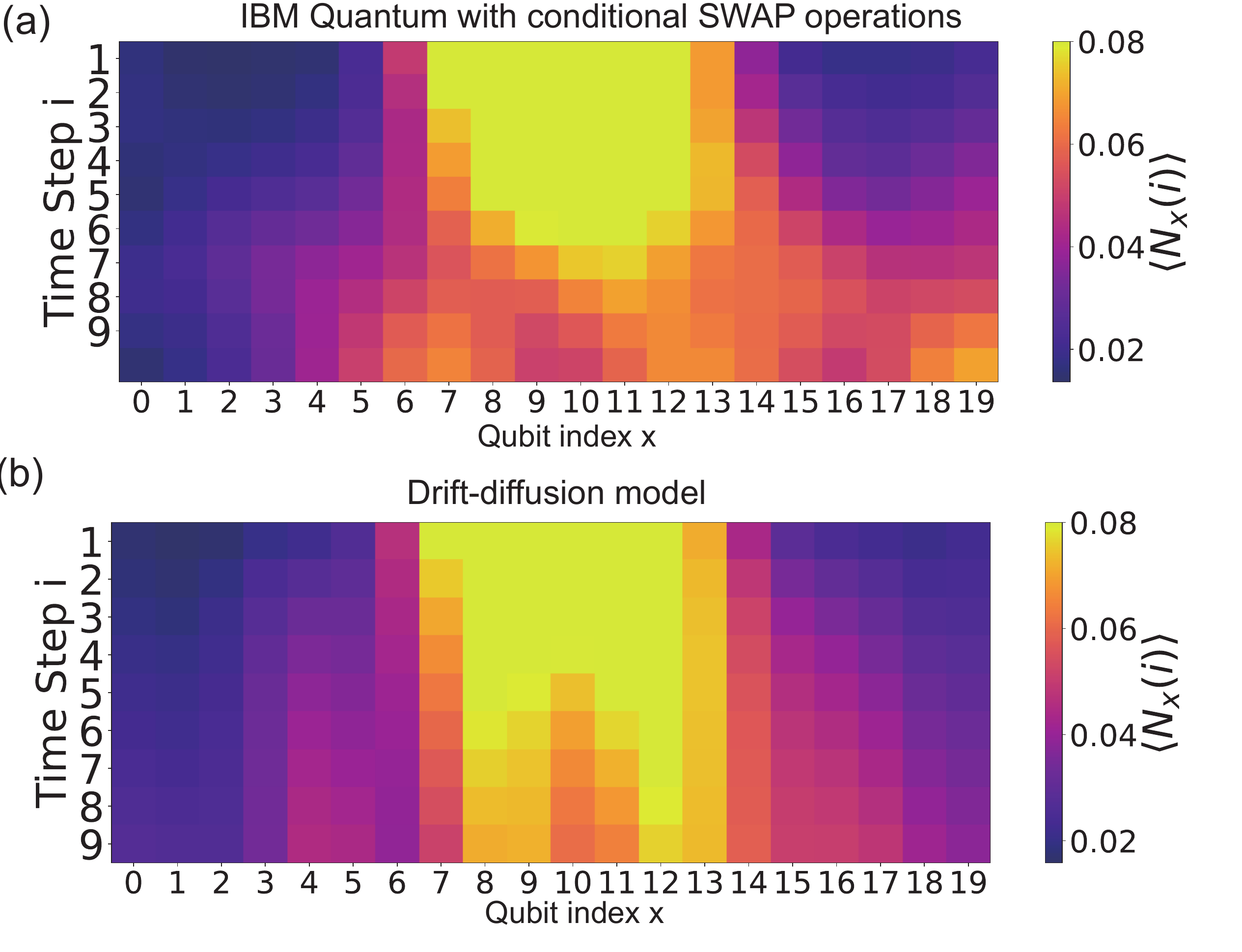}
    \caption{(a) Measured space–time density profile obtained from IBM Quantum hardware, showing the directed evolution of local occupations under the conditional-SWAP feedback protocol. These data are adopted from Fig.2 (main text), and the density is normalized as $N_{x}(i)\rightarrow N_{x}(i)/\sum_{x} N_{x}(i)$. (b) Drift--diffusion reconstruction of the same dynamics, obtained by fitting the measured data to the higher-order continuum model in Eq.~\ref{pdedirft}, with fitted parameters $v=0.1024768$ and $[D_{2},D_{3},D_{4}]=[1.21687893,-0.04005582,1.2844856]$. Here, the reconstructed profile shows the asymmetric flow trend.}
    \label{fig:classicaldata2}
\end{figure}

\section*{Supplementary Note 6. Additional discussion on the circuit scalability}

To assess scalability in a hardware-realistic way, we estimate the dominant time and error budgets for the IBM backend used in the large-scale experiments. Here, the quoted coherence times, gate durations, and error rates are taken from the backend calibration snapshot used for the corresponding experiment, and we use the reported median values over the selected qubits and couplers on the device. Concretely, we use
\begin{equation}
T_1 \simeq 142.5~\mu{\rm s},
\qquad
T_2 \simeq 95.1~\mu{\rm s},
\end{equation}
together with a median single-qubit gate duration
\begin{equation}
\tau_{1q}\simeq 24~{\rm ns},
\end{equation}
a median two-qubit CZ-gate duration and error
\begin{equation}
\tau_{2q}\simeq 68~{\rm ns},
\qquad
r_{2q}\simeq 2.6\times10^{-3},
\end{equation}
and a mid-circuit measurement duration and assignment error
\begin{equation}
\tau_m\simeq 1.56~\mu{\rm s},
\qquad
r_m\simeq 9.4\times10^{-3}.
\end{equation}

Our scrambling layer is compiled as two parallel CZ sublayers (odd and even bonds) together with two single-qubit rotation layers. Its unitary duration is therefore estimated as
\begin{equation}
\tau_U \approx 2\tau_{2q}+2\tau_{1q}
\simeq 184~{\rm ns}.
\label{eq:tauU}
\end{equation}
For a feedback layer that contains one mid-circuit measurement round followed by a conditional single-qubit operation, the corresponding duration is
\begin{equation}
\tau_{\rm layer}
\approx
\tau_U+\tau_m+\tau_{\rm cond},
\label{eq:tau_layer}
\end{equation}
where $\tau_{\rm cond}$ denotes the duration of the conditional pulse. For a conditional $X$ operation, it takes
\begin{equation}
\tau_{\rm cond}\sim \tau_{1q},
\end{equation}
so that
\begin{equation}
\tau_{\rm layer}
\approx
184~{\rm ns}+1.56~\mu{\rm s}+24~{\rm ns}
=
1.768~\mu{\rm s}.
\end{equation}
Since $10\,\tau_{\rm layer}\approx 17.68~\mu{\rm s}$
is still substantially smaller than the measured dephasing time $T_2\simeq 95.1~\mu{\rm s}$, the $\sim 10$-layer circuits studied in the main text lie within a reasonable coherence budget. This supports the feasibility of the large-scale hardware implementation.

For a circuit of width $L$ and depth $n$, we estimate the survival fidelity by the conservative hardware-level proxy
\begin{equation}
	F(n,L)\approx \exp\!\left[-n\,\varepsilon_{\rm layer}^{\rm eff}(L)\right].
\end{equation}
For a brickwork nearest-neighbour scrambling layer on an open chain, the number of two-qubit CZ gates in one full layer is $N_{2q}(L)
=
\left\lfloor \frac{L}{2}\right\rfloor
+
\left\lfloor \frac{L-1}{2}\right\rfloor
=
L-1$. 
Accordingly, the effective error accumulated in a single layer may be estimated as
\begin{equation}
	\varepsilon_{\rm layer}^{\rm eff}(L)
	\approx
	N_{2q}(L)\,r_{2q}
	+
	N_m(L)\,r_m
	+
	\frac{\tau_{\rm layer}}{T_2},
	\label{eq:eps_scaling_compare}
\end{equation}
where $r_{2q}$ is the error per two-qubit gate, $r_m$ is the mid-circuit measurement error, $N_m(L)$ is the expected number of mid-circuit measurements executed in one layer, and $\tau_{\rm layer}$ is the wall-clock duration of a full layer, including the measurement-feedback block. Since the CZ gates within each even or odd sublayer are executed in parallel, $\tau_{\rm layer}$ does not grow extensively with $L$, whereas the stochastic gate-error contribution does.

In the sparse-feedback regime, the expected number of mid-circuit measurements per layer is approximately $N_m(L)\approx pL$, so Eq.~\eqref{eq:eps_scaling_compare} reduces to
\begin{equation}
	\varepsilon_{\rm layer}^{\rm eff}(L)
	\approx
	(L-1)\,r_{2q}
	+
	pL\,r_m
	+
	\frac{\tau_{\rm layer}}{T_2}.
	\label{eq:eps_nL}
\end{equation}

For a sparse-feedback setting with $p=0.01$, this gives
\begin{align}
	\varepsilon_{\rm layer}^{\rm eff}(50)
	&\approx
	49\times 2.6\times10^{-3}
	+
	0.5\times 9.4\times10^{-3}
	+
	\frac{1.768}{95.1}
	\simeq
	1.51\times10^{-1},
	\\
	\varepsilon_{\rm layer}^{\rm eff}(80)
	&\approx
	79\times 2.6\times10^{-3}
	+
	0.8\times 9.4\times10^{-3}
	+
	\frac{1.768}{95.1}
	\simeq
	2.32\times10^{-1},
	\\
	\varepsilon_{\rm layer}^{\rm eff}(100)
	&\approx
	99\times 2.6\times10^{-3}
	+
	1.0\times 9.4\times10^{-3}
	+
	\frac{1.768}{95.1}
	\simeq
	2.85\times10^{-1}.
	\label{eq:eps_numeric}
\end{align}

If we instead consider a denser-feedback regime with $p=0.1$, the per-layer error increases further:
\begin{align}
	\varepsilon_{\rm layer}^{\rm eff}(50)
	&\approx
	49\times 2.6\times10^{-3}
	+
	5\times 9.4\times10^{-3}
	+
	\frac{1.768}{95.1}
	\simeq
	1.93\times10^{-1},
	\\
	\varepsilon_{\rm layer}^{\rm eff}(80)
	&\approx
	79\times 2.6\times10^{-3}
	+
	8\times 9.4\times10^{-3}
	+
	\frac{1.768}{95.1}
	\simeq
	2.99\times10^{-1},
	\\
	\varepsilon_{\rm layer}^{\rm eff}(100)
	&\approx
	99\times 2.6\times10^{-3}
	+
	10\times 9.4\times10^{-3}
	+
	\frac{1.768}{95.1}
	\simeq
	3.70\times10^{-1}.
	\label{eq:eps_numeric_p01}
\end{align}

These estimates show that increasing the feedback rate substantially amplifies the layer error, especially at large system size. For this reason, the main-text experiments are carried out in the low-measurement regime. In that regime, the additional error coming from mid-circuit measurement remains comparatively controlled, allowing the feedback-induced transport signatures to remain observable  at the 100-qubit scale.

\flushbottom

\end{document}